\documentclass[12pt]{article}
\usepackage{amssymb,amsmath}
\usepackage{bm} 
\usepackage{booktabs} 
\usepackage{array}
\usepackage{latexsym}
\usepackage{graphicx}
\usepackage{color}
\usepackage{datetime}
\usepackage[nosort]{cite}
\usepackage{verbatim}
\usepackage{enumerate}
\usepackage{chngpage} 
\usepackage{mathrsfs}
\usepackage{euscript}
\usepackage{psfrag}
\usepackage{subfig}
\usepackage{anyfontsize}
\usepackage{braket}
\usepackage{subcaption}
\usepackage{blindtext}%
\usepackage[theorems,breakable]{tcolorbox}%
\usepackage{mathtools}  
\usepackage{xfrac}  

\usepackage{multirow,graphicx}
\usepackage{makecell}
\usepackage{float}
\usepackage{tikz}
\usetikzlibrary{decorations.text}
\usepackage{enumitem}
\setlist{itemsep=0pt}
\usetikzlibrary{math}
\usepackage{setspace}
\usepackage{cancel}

\usepackage[latin1]{inputenc}
\usepackage{aurical}
\usepackage{anyfontsize}

\usepackage{mciteplus}

\usepackage[colorlinks=true,      linkcolor=blue,      urlcolor=blue,      
            filecolor=blue,      citecolor=blue,       pdfstartview=FitH,     
						pdfpagemode=UseNone,      bookmarksopen=true]{hyperref}  
\usepackage[all]{hypcap}     

\usepackage{tikz}
\usepackage{pgfplots}
\usetikzlibrary{decorations.text,intersections, pgfplots.fillbetween}
\usetikzlibrary{math}
 \usetikzlibrary{3d,shapes.geometric,shadows.blur}
\usepackage{tikz-3dplot}

\usepackage{blindtext}%
\usepackage[theorems,breakable]{tcolorbox}%

\newtcbtheorem{mytheo}{}%
{colback=gray!5,colframe=gray!35!black,fonttitle=\bfseries}{th}

\newtcbtheorem{lem}{}{%
        colback=green!5,%
        colframe=green!35!black,%
        fonttitle=\bfseries,title %
    }{lem}

\definecolor{amaranthred}{rgb}{0.83,0.13,0.18}
\definecolor{amazon}{rgb}{0.23,0.48,0.34}
\definecolor{bdazzledblue}{rgb}{0.18,0.35,0.58}
\definecolor{absolutezero}{rgb}{0.0,0.28,0.73}
\definecolor{bitterlemon}{rgb}{0.79,0.88,0.05}
\definecolor{byzantine}{rgb}{0.74,0.2,0.64}
\definecolor{turquoise}{rgb}{0.19, 0.84, 0.78}
\definecolor{burgundy}{rgb}{0.5, 0.0, 0.13}
\definecolor{airforceblue}{rgb}{0.36, 0.54, 0.66}
\definecolor{arsenic}{rgb}{0.23, 0.27, 0.29}

\newcommand{\comm}[1]{} 

\def\({\left(}
\def\){\right)}
\def\[{\left[}
\def\]{\right]}

\def\One{{\hbox{ 1\kern-.8mm l}}}

\def\barray{\begin{array}}
\def\earray{\end{array}}
\def\be{\begin{equation}}
\def\ee{\end{equation}}
\def\bea{\begin{eqnarray}}
\def\eea{\end{eqnarray}}
\def\bal{\begin{align}}
\def\eal{\end{align}}
\def\nn{\nonumber}

\def\-{\,-\,}
\def\={\,=\,}
\def\+{\,+\,}
\def\equi{\,\equiv\,}

\def\cK{{\cal K}}

\numberwithin{equation}{section} 

\definecolor{cardinal}{rgb}{0.6,0,0}
\definecolor{darkgreen}{rgb}{0,0.4,0}
\definecolor{golden}{rgb}{0.92, 0.7, 0}
\definecolor{midnight}{rgb}{0, 0, 0.5}
\definecolor{darkblue}{rgb}{0, 0, 0.7}
\definecolor{purple}{rgb}{0.5, 0, 0.5}

\definecolor{amaranthred}{rgb}{0.83,0.13,0.18}
\definecolor{amazon}{rgb}{0.23,0.48,0.34}
\definecolor{bdazzledblue}{rgb}{0.18,0.35,0.58}
\definecolor{absolutezero}{rgb}{0.0,0.28,0.73}
\definecolor{bitterlemon}{rgb}{0.79,0.88,0.05}
\definecolor{byzantine}{rgb}{0.74,0.2,0.64}
\definecolor{turquoise}{rgb}{0.19, 0.84, 0.78}
\definecolor{burgundy}{rgb}{0.5, 0.0, 0.13}

\def\cA{{\cal A}}

\def\cE{{\cal E}}

\def\cL{{\cal L}}
\def\cM{{\cal M}}
\def\cN{{\cal N}}

\def\cO{{\cal O}}

\pgfplotsset{compat=1.18}
\begin{document}

\begin{titlepage}
\thispagestyle{empty}

\begin{flushright}

\end{flushright}

\begin{center}
\noindent{\bf \Large Localized Black Holes in AdS$_3$:  }\\
\vspace{0.2cm}
\noindent{\bf \Large What Happens Below $\sfrac{c}{12}$ Stays Below $\sfrac{c}{12}$}\\
\vspace{0.2cm}

\vspace{1cm}

{\bf \normalsize Iosif Bena$^a$, Rapha\"el Dulac$^a$, Pierre Heidmann$^b$, and Zixia Wei$^{c, d}$}
\vspace{1cm}\\

${}^a${\it
Institut de Physique Th\'eorique,
Universit\'e Paris Saclay, 
CEA, CNRS, 
Orme des Merisiers, Gif sur Yvette, 91191 CEDEX, France
}\\[1.5mm]

${}^b${\it
Department of Physics and Center for Cosmology and AstroParticle Physics (CCAPP), \\
The Ohio State University, Columbus, OH 43210, USA
}\\[1.5mm]

${}^c${\it
Jefferson Physical Laboratory,
Harvard University, Cambridge, MA 02138, USA
}\\[1.5mm]

${}^d${\it
Society of Fellows,
Harvard University, Cambridge, MA 02138, USA
}\\[1.5mm]

\end{center}

\begin{abstract}

We construct asymptotically AdS$_3 \times$S$^3 \times$T$^4$ black holes that are localized on the S$^3$ and co-exist with the BTZ black hole at small positive energies. These black holes dominate the microcanonical ensemble for $ E\leq \frac{c}{24}\left(5\sqrt{5}-11\right)$, suggesting they could represent the endpoint of the BTZ instability at low energies. Remarkably, they also exist at negative energies, where pure Einstein gravity predicts no states and the BTZ black hole does not exist. They appear in the spectrum immediately above $-\frac{c}{12}$ (the energy of global AdS$_3$), and their entropy is a significant fraction (up to 1/2) of the entropy of the free orbifold CFT at negative energies. Our solutions exist in an energy window outside the universal predictable range of the modular bootstrap in large-$c$ CFT$_2$ and, despite their microcanonical dominance, do not dominate in the canonical ensemble.

To calculate the holographic entanglement entropy of our solutions, we propose the first recipe that can be applied to arbitrary geometries asymptotic to AdS$_3$ times an internal manifold, and depend non-trivially on its coordinates. We find that our new geometries have an entanglement entropy nearly identical to that of the BTZ black hole with the same energy, despite having different horizon structures. However, they can be distinguished by non-minimal extremal surfaces, which unveil finer details of the microstructure.

\end{abstract}

\end{titlepage}

\newpage
\setcounter{page}{1}
\renewcommand{\baselinestretch}{1.2}\normalsize
\tableofcontents
\renewcommand{\baselinestretch}{1.24}\normalsize

\section{Introduction}
\label{sec:intro}

Black holes with $\rm AdS_3$ asymptotics encode the thermodynamic properties of families of states in the dual 1+1-dimensional CFT through the AdS/CFT correspondence \cite{Maldacena:1997re,GKP98,Witten98}. Since these CFTs are among the most extensively studied strongly coupled theories, their physics can be compared with the insights these black holes provide. Such comparisons can refine the holographic correspondence and illuminate black hole physics.

The working horse of asymptotically-$\rm AdS_3$ black holes is the BTZ black hole \cite{Banados:1992wn}, which captures the thermodynamic properties of the dual CFT at high energies. However, as the energy decreases to the range $-c/12 < E < 0$, this black hole ceases to exist, leaving the thermodynamic properties of the CFT in this regime more enigmatic.\footnote{Here, $-c/12$ represents the energy of global $\rm AdS_3$, corresponding to the ground state of the dual $\rm CFT_2$, and $c$ is its central charge.}

For pure Einstein gravity in three dimensions, there are no primary states in this energy interval. By contrast, symmetric orbifold CFTs exhibit a significant degeneracy of states, with entropy of order $c$ \cite{Hartman:2014oaa}.  Very little is known about the fate of these states as one transitions from weak to strong coupling, particularly whether all these CFT states are lifted, as suggested by pure Einstein gravity, or whether some persist and admit a semi-classical gravitational description. Modular bootstrap arguments impose only a loose constraint on the spectrum of states in generic large-$c$ CFTs within the so-called ``medium and low energy range," $-c/12\leq E \leq c/12$ \cite{Hartman:2014oaa}. This constraint is called the ``sparseness" condition, requiring the spectrum to be relatively sparse with an entropy as a function of energy smaller than $$S(E) \,\lesssim\, 2\pi\left(E+\frac{c}{12}\right)\,.$$
This condition is saturated in symmetric-orbifold CFTs.

Nevertheless, a few hints provide guidance. When embedded String Theory as a solution with  $\rm AdS_3 \times S^3 \times T^4$ asymptotics, the BTZ black hole preserves the isometries of the ${\rm S}^3$, and can be viewed as the ``large'' black hole. As the energy decreases, it is expected that this black hole undergoes a Gregory-Laflamme-type instability \cite{Banks:1998dd,Peet:1998cr,
Gregory:1993vy} towards a ``small'' black hole, localized on the ${\rm S}^3$, similar to what happens for $\rm AdS_5 \times S^5$ large black holes \cite{Hubeny:2002xn,Buchel:2015gxa,Dias:2016eto}. For supersymmetric spinning BTZ black holes, which descend from the dimensional reduction of BMPV black holes   \cite{Breckenridge:1996is}, one can similarly construct solutions localized on the ${\rm S}^3$ that exist outside the BTZ parameter space \cite{Gauntlett:2004qy, Bena:2011zw}. Remarkably, the entropy of these solutions can dominate over the BTZ entropy in a narrow range. \\

This paper aims to explore the fate of non-supersymmetric states in AdS$_3$/CFT$_2$ at medium and low energies, $-c/12\leq E \leq c/12$. We construct families of static asymptotically ${\rm AdS}_3 \times {\rm S}^3 \times {\rm T}^4$  solutions that describe black holes localized on the $\rm S^3$. These black holes are analytic solutions derived in \cite{Bah:2022pdn} using a novel integrable structure for static cohomogeneity-two solutions in supergravity \cite{Heidmann:2021cms,Bah:2021owp,Bah:2021rki,Heidmann:2022zyd,Bah:2022yji,Bah:2023ows,Heidmann:2023kry}.

We show that these static localized black holes exist in the energy interval, $-c/12 < E \leq c/96$, populating the spectrum in the low and medium energy range. They have an entropy of order $c$, larger than the entropy of the static BTZ black hole when $0\leq E<\frac{c}{24}\left(5\sqrt{5}-11\right)\sim \frac{3c}{400}$. Hence, these black holes dominate the microcanonical ensemble in this energy range. Furthermore, their entropies satisfy the CFT sparseness bound \cite{Hartman:2014oaa}, and account for a finite fraction of the entropy of the D1-D5 CFT at the free symmetric orbifold point.\footnote{This fraction is exactly $\frac{1}{2}$ at $E=0$. A similar phenomenon happens for non-supersymmetric states of the $\cN=4$ SYM theory, whose strong-coupling entropy is $3/4$ of the weak-coupling one\cite{Gubser:1996de,Klebanov:2000me}.} Thus, a large number of states exist in the range $-c/12 < E < 0$, both at weak and strong coupling, contrary to naive expectations based on pure Einstein gravity.

The existence of these solutions may appear surprising given the universal constraints derived from the modular invariance of large-$c$ CFTs, which suggest that the BTZ black hole captures entirely the canonical-ensemble physics \cite{Hartman:2014oaa}. We demonstrate that, despite the microcanonical dominance of localized black holes, the BTZ black hole remains the most dominant saddle in the canonical ensemble. The dominance of our new solutions occurs inside what we call the \emph{canonical shadow}, the energy interval where the universal constraints of large-$c$ CFTs fail to capture the microcanonical-ensemble physics. 

In the microcanonical ensemble, we show that (within our isometry class) the BTZ black hole undergoes a phase transition at $E=\frac{c}{24}\left(5\sqrt{5}-11\right)$, leading to the formation of a localized black hole geometry. This occurs because the energy of the BTZ black hole becomes insufficient to sustain a horizon spread over the entire S$^3$. As a result, an instability changes the topology of the solution, creating a horizon covering a smaller region of the S$^3$: a \emph{black pole}. 

Remarkably, at the transition point, the black pole has a S$^3$ split into two regions, with the horizon confined to $0\leq \cos^2 \theta \leq \varphi^{-1}$, where $\theta$ is the angular position in the S$^3$ Hopf coordinates and $\varphi$ is the golden ratio. At even lower energies, one might expect further fragmentation of the black region into multiple smaller black holes. However, we show that the black pole remains dominant in the microcanonical ensemble, with its horizon shrinking in size as the energy decreases. At $E\sim -c/12$, the geometry reduces to being predominantly global AdS$_3$, with an infinitesimal black hole localized at the pole of the S$^3$ and the center of AdS$_3$. \\

An important question is whether signs of this phase transition can be captured in the dual CFT.\footnote{For supersymmetric black holes, phases with more entropy than the BTZ black hole can be identified both in the boundary theory and the bulk \cite{Bena:2011zw}.} 
Naturally, since the black pole breaks the $SO(4)$ internal symmetry corresponding to $\rm S^3$ rotations into a $U(1)\times U(1)$ subgroup, certain operators in the dual CFT acquire nonzero vacuum expectation values. However, this provides limited insight into the microscopic structure of the states on the two sides of the transition. 

A natural quantity that can reflect microscopic details of different black hole geometries is the entanglement entropy. As argued in \cite{Hayden:2020vyo,Wei:2022cpj}, an exponentially large number of ``disentangled" states in the CFT exhibit entanglement entropies that differ significantly from those of the typical black hole microstates. Since the solutions we construct break the rotational symmetry of the BTZ horizon, it is possible they might have an entanglement entropy parametrically different from that of the BTZ black hole and turn out to offer a statistical ensemble description of some disentangled states in gravity.

To compute holographically the entanglement entropy of a subsystem in the CFT, one applies the Ryu-Takayanagi (RT) procedure \cite{RT06,RT06b}. This involves finding a minimal surface (the minimal RT surface) that ends on the boundaries of the subsystem and extends into the bulk geometry. For AdS$_3$, the boundary is a circle that can be normalized to a length of $2\pi$, allowing any connected boundary subsystem to be represented as an interval characterized by its length $0\leq L\leq \pi$. 

For simple product geometries, $\rm AdS_3 \times \mathcal{M}$, finding the RT surface reduces to a minimization problem, which has been extensively studied \cite{RT06,RT06b,RT17}. In contrast, determining the RT surface in more complex geometries lacking a product structure poses significant challenges. Solutions have only been obtained in restricted cases, such as warped product geometries that preserve the internal-space isometries \cite{Bombini:2019vuk}, or for RT surfaces constrained near the AdS boundary, avoiding the deeper, more intricate regions of the bulk \cite{Giusto:2014aba}.

We formulate a novel method to estimate the area of RT surfaces in static geometries with nontrivial dependence on the internal directions. To our knowledge, this is the first such calculation performed in solutions where the metric along the AdS directions exhibits significant dependence on the internal directions, and we believe our calculations can be generalized in other similarly complicated $\rm AdS \times \mathcal{M}$ backgrounds. \\

{Our calculations reveal an intriguing result: although the localized black hole geometries differ significantly from the BTZ black hole in the infrared, they have the same entanglement entropy (up to small corrections of order $E/c$). Such a match is similar to what happens for small values of $L$ where the minimal RT surfaces remain close to the AdS boundary, and the entanglement entropy obeys a universal behavior because of the entanglement first law \cite{RT17,BNTU12,WKZV13,BCHM13,FGHMvR13,TT21}. Surprisingly, we find that for the localized black holes this
match persists for large intervals, up to $L=\pi$, with the entanglement entropy depending solely on macroscopic quantities, such as the central charge and energy, and remaining independent of the underlying microstructure.

This happens because our solutions only exist in the energy range, $E<c/96$. The deviations of the BTZ entanglement entropy from the universal behavior are in fact suppressed by corrections of order $3E/c$, even for large $L$. One could therefore argue that other states at very low energies will have an entanglement entropy that approaches the universal behavior as $E/c \rightarrow 0$. }

From a bulk perspective, this occurs because our geometries deviate from the BTZ black hole only deep in the infrared, within the so-called \emph{entanglement shadow.} This is the region in AdS that is inaccessible to minimal RT surfaces \cite{Balasubramanian:2014sra}.
Consequently, the BTZ black hole and the localized black holes can be distinguished only through holographic observables that probe the entanglement shadow. As proposed in \cite{Balasubramanian:2014sra,Caminiti:2024ctd,Arora:2024edk}, extremal but non-minimal RT surfaces are expected to capture internal degrees of freedom within the entanglement shadow. In the dual CFT, this is related to a nontrivial correlation between specific subsets of the internal degrees of freedom \cite{Balasubramanian:2014sra}. These surfaces are direct continuations of the minimal RT surfaces of the $(0,\pi)$ interval to intervals larger than $\pi$. Their area grows with the interval length as the geodesic loops around within the entanglement shadow (refer to Fig.\ref{fig:EntangEntwin} for a schematic description).

In black-hole geometries for which the horizon is confined within their entanglement shadow, non-minimal RT surfaces orbit increasingly close to the horizon, causing their area to exhibit a linear growth with the length, proportional to the black-hole entropy. This behavior encodes microscopic details about the states. Since localized black holes have larger entropy than BTZ black holes at low energy, we show that non-minimal RT surfaces provide the holographic observables that differentiate between these solutions. 

\medskip

The localized black holes ``live in the shadows'' in two important ways. First, they dominate the microcanonical ensemble within an energy range that is entirely invisible to the canonical ensemble. Second, their geometric differences from the BTZ black hole are confined to the  entanglement shadow of the later. Hence these differences are barely visible through the calculation of the entanglement entropy. As a result, these geometries occupy the \emph{canonical shadow}, as well as the \emph{entanglement shadow}. To capture their physics, more refined observables sensitive to microscopic details of the internal degrees of freedom are required, such as non-minimal RT surfaces.

{The alignment of the canonical shadow and the entanglement shadow is striking, and arises from several highly nontrivial facts. First, none of our solutions exist in the range $c/96<E<c/12$. Within this range, the BTZ entanglement entropy for finite intervals starts to differ significantly from the universal formula, reflecting infrared details of the geometry. If other solutions existed in this energy regime, their entanglement entropy would likely differ both from the universal formula and from that of the BTZ black hole. }

{Second, one could attempt to construct solutions that ``escape'' the entanglement shadow, by adding angular momentum that increases their size. However, this would reduce their entropy, making them less likely to dominate the microcanonical ensemble.} These facts lead us to put forth the following\vspace{0.2cm}\\
{\bf Shadow Conjecture}: \\
{\em A black hole that dominates the large AdS black hole in the microcanonical ensemble must differ from it only within its entanglement shadow.} \vspace{0.2cm}

It would be highly valuable to test the validity of this conjecture on other lumpy solutions that dominate the microcanonical ensemble but are not visible in the canonical ensemble \cite{Dias:2016eto,Dias:2015pda,Dias:2017uyv,Dias:2024vsc,Herzog:2017qwp}, to construct additional families of such solutions \cite{Dias-Santos-toappear} and try to provide additional arguments for this conjecture using holographic CFT or bootstrap methods. We believe the conjecture can also be tested for certain families of supersymmetric black holes with AdS asymptotics \cite{Gauntlett:2004wh,Denef:2007vg, deBoer:2008fk,Bena:2011zw}.

In Section \ref{sec:2}, we review key results from modular bootstrap in large-$c$ CFTs and from pure Einstein gravity in three dimensions. In Section \ref{Sec: Sugra solution}, we present a new family of localized black holes that break the isometries of the BTZ horizon and exist in the energy range $-c/12<E<c/96$. In Section \ref{sec:ThermoLowEn}, we derive their thermodynamics at $E<0$, demonstrating that these black holes account for a significant fraction of the entropy of the D1-D5 free-orbifold CFT within this energy range. In Section \ref{sec:PhaseDiag}, we analyze their thermodynamic behavior at positive energies, above the BTZ threshold, and show that while they dominate the microcanonical ensemble within a specific energy range, this dominance is not reflected in the canonical ensemble. In Section \ref{sec:EEGen}, we introduce a novel method for estimating minimal RT surfaces in static geometries where the metric along the AdS coordinates depends nontrivially on the internal directions, and we apply it to the localized black holes. Conclusions and future directions are discussed in Section \ref{sec:Conclusion}, while additional details regarding the entanglement entropy calculations are provided in Appendix \ref{app:1} and \ref{app:pure_mixed}.

Readers primarily interested in the derivation of the entanglement entropy in our nontrivially-fibered solutions are encouraged to focus on Section \ref{sec:EEGen}, while those interested in the supergravity solutions and their thermodynamic properties may find Sections \ref{Sec: Sugra solution}, \ref{sec:ThermoLowEn} and \ref{sec:PhaseDiag} most relevant.

\section{CFT expectations and pure Einstein gravity}

\label{sec:2}

In this section, we briefly review established results on AdS$_3$/CFT$_2$ holography. We begin by summarizing universal thermodynamic properties and the energy spectrum from a modular bootstrap perspective, followed by their manifestations in three-dimensional pure Einstein gravity.

\subsection{Statistical ensembles in large-\texorpdfstring{$c$}{c} CFT}
\label{sec:CFT}

We focus on universal features in the energy spectrum of a two-dimensional CFT at large central charge, $c \rightarrow \infty$, which can be inferred from modular invariance. In \cite{Hartman:2014oaa}, it was shown that the simple assumption that the low-energy spectrum is \emph{sparse}:
\begin{align}\label{eq:sparse_spectrum}
    S(E) \,\lesssim\, 2\pi\left(E+\frac{c}{12}\right),\qquad (E\leq 0), 
\end{align}
where $S$ is the entropy and $E$ the energy, imposes strong constraints on the canonical and microcanonical ensembles. Note that we have assumed that the spatial direction of the two-dimensional CFT has a period of $2\pi$.

\subsubsection{Universality of canonical ensembles}

\begin{figure}[t]
    \centering
    \includegraphics[width=\textwidth]{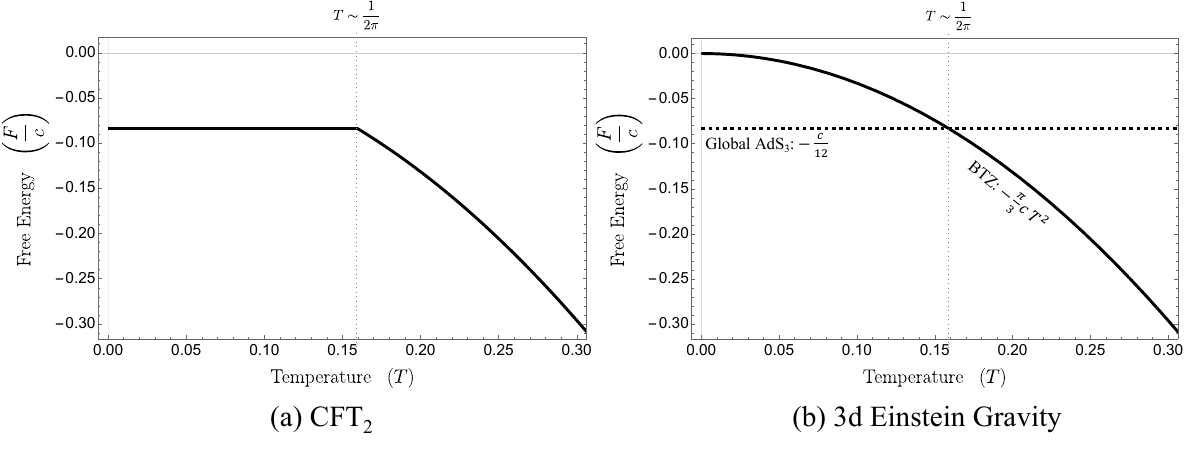}
    \caption{Properties of the canonical ensembles. Left plot: The free energy as a function of the temperature, $T$, in large-$c$ CFT$_2$, as predicted by modular bootstrap. Right plot: The free energy of thermal AdS$_3$ and the BTZ black hole in pure AdS$_3$ Einstein gravity. There is a first-order phase transition at $T = (2\pi)^{-1}$ between the two saddle points. The minimal free energy in AdS$_3$ Einstein gravity matches the universal result in CFT$_2$.}
    \label{fig:CFTbootstrap_F}
\end{figure}

In \cite{Hartman:2014oaa}, it has been shown that the sparseness condition \eqref{eq:sparse_spectrum} imposes a universal dependence of the free energy on the temperature, $T$,
\begin{align}
    F(T) \= 
    \left\{
\begin{array}{ll}
-\frac{c}{12} & \text{when } T\leq \frac{1}{2\pi}\,,\\
-\frac{c}{12} \left(2\pi T\right)^2&\text{when } T> \frac{1}{2\pi}\,,
\end{array}
\right. 
\end{align}
depicted on the left plot in Fig.\ref{fig:CFTbootstrap_F}.

Thus, the \emph{behavior of canonical ensembles becomes universal} and fully determined in the large-$c$ limit. 

\subsubsection{Microcanonical ensembles and the canonical shadow}

The entropy and energy at fixed temperature are derived from the free energy:\footnote{We remind the reader that $S(T)=-\frac{\partial F}{\partial T}$ and $E(T)=F(T)+ T S(T)$.}
\begin{align}\label{eq:entropy_beta}
    S(T) = 
    \left\{
\begin{array}{ll}
\cO(1) & \text{when } T\leq \frac{1}{2\pi}\,,\\
\frac{c}{6}(2\pi)^2 T & \text{when } T> \frac{1}{2\pi}\,,
\end{array}
\right.  
\quad 
E(T)  = 
\left\{
\begin{array}{ll}
-\frac{c}{12}  &  \text{when } T\leq \frac{1}{2\pi}\,,\\
\frac{c}{12}(2\pi T)^2  & \text{when } T> \frac{1}{2\pi}\,.
\end{array}
\right.
\end{align}
Remarkably, the energy jumps from $-c/12$ to $c/12$ at the phase transition $T = (2\pi)^{-1}$. This suggests a ``shadow" in the energy spectrum within this range, which remains undetermined by universal large-$c$ CFT constraints.

The equivalence of statistical ensembles,\footnote{The equivalence of statistical ensembles requires the entropy to be a concave function of energy \cite{gross2001microcanonical}. However, as we will see, $S(E)$ is not concave for the systems considered in this paper, which is a key factor in the mismatch between the canonical and microcanonical ensembles.} indicates that the behavior of the microcanonical ensembles (at fixed energy $E$) is largely determined, except in the energy range $-c/12 \leq E \leq c/12$, which is not well captured by the canonical ensemble. 

Accordingly, we can define three energy ranges for the microcanonical ensemble: 
\begin{itemize}
    \item \underline{Low energies:} $-\frac{c}{12} \leq E \leq 0$.
    \item \underline{Medium energies:} $0 \leq E \leq \frac{c}{12}$.
    \item \underline{High energies:} $\frac{c}{12} \leq E$.
\end{itemize}
The properties of the microcanonical ensemble in the high-energy range are expected (from generic CFT arguments) to be universally determined by the entropy function\footnote{The entropy at high energy is given by \eqref{eq:entropy_beta}, replacing $T$ with $E$ for $T>(2\pi)^{-1}$.}
\begin{align}
    S(E) = 2\pi \sqrt{\frac{c}{3} E} ,\qquad \text{when } E\geq \frac{c}{12}\,,
\end{align}
On the other hand, the entropy in the low- and medium-energy ranges cannot be universally determined but admits loose bounds from the sparseness condition \eqref{eq:sparse_spectrum}\footnote{Note that the upper bound for the medium energy range is proven, while that for the low energy range is assumed. Also note that the lower bound for the medium energy range is not proven, but it is expected to hold generically as breaking it needs fine-tuning in the CFT spectrum \cite{Hartman:2014oaa}.}
\begin{align}\label{eq:medium_up}
   S_{min} \,\lesssim\, S(E) \,\lesssim\, 2\pi\left(E+\frac{c}{12}\right) ,\qquad \text{when } -\frac{c}{12}\leq E\leq \frac{c}{12}\,,
\end{align}
where $S_{min}=0$ for the low-energy range and $S_{min}=2\pi \sqrt{\frac{c}{3}E}$ at medium energies.

In \cite{Hartman:2014oaa}, it was shown that symmetric orbifold CFTs saturate the entropy upper bound, indicating a significant degeneracy of CFT states in the low- and medium-energy range. This includes the well-known D1-D5 CFT at the orbifold point, which is dual to type String Theory on $\rm AdS_3 \times S^3 \times T^4 $. However, the fate of these states is not known in the regime of parameters where the $\rm AdS_3$ radius is large (strong coupling). 
Anticipating the results of the subsequent sections, we will construct black hole geometries that account for a significant portion of the entropy of these states in these regimes, demonstrating that a substantial number of them are not lifted.

\begin{figure}[t]
    \centering
    \includegraphics[width= \textwidth]{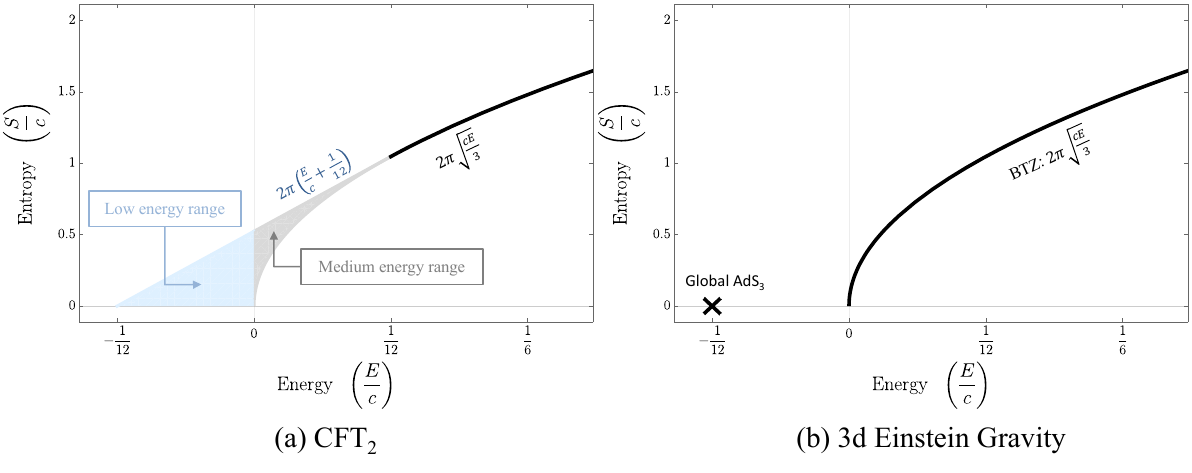}
    \caption{Properties of the microcanonical ensembles.
    Left plot: The entropy as a function of energy, $S(E)$, in large-$c$ CFT$_2$ predicted by modular bootstrap. The blue and gray regions are the  loose CFT bounds from the sparseness condition \eqref{eq:sparse_spectrum} (saturated by symmetric-orbifold CFTs). The black curve shows the universal behavior in the high-energy range. Right plot: The entropy as a function of the energy in pure AdS$_3$ Einstein gravity. Beyond the black hole threshold $E > 0$, there is a continuous spectrum of the BTZ black hole microstates. In the low-energy range, the only on-shell configuration is global AdS$_3$ at $E = -c/12$, which is depicted as $\times$ in the figure. We can see that the entropy in AdS$_3$ Einstein gravity is consistent with the large-$c$ CFT$_2$ result. }
    \label{fig:CFTbootstrap}
\end{figure}

The left plot in Fig.\ref{fig:CFTbootstrap} illustrates the universal CFT expectations for the microcanonical ensembles. The shaded regions indicate the canonical shadow of CFT uncertainty in the low and medium energy spectrum.

\subsection{Thermodynamics in \texorpdfstring{AdS$_3$}{AdS3} Einstein gravity}
\label{sec:EinsteinGrav}

According to the AdS/CFT correspondence, the semiclassical limit of an AdS$_3$ gravitational theory (possibly fibered over a compact space) should be dual to some large-$c$ CFT$_2$, and be consistent with the large-$c$ CFT$_2$ predictions reviewed previously.

As a foundational example, consider pure Einstein gravity in AdS$_3$ with action:
\begin{align}
    I_{\rm grav.} = -\frac{1}{16\pi G_3}\int_{\mathcal{M}} \sqrt{g} \left(R-\frac{2}{R_{AdS}^2}\right)+ ({\text{Gibbons-Hawking term}}), 
\end{align}
where $R_{AdS}$ is the AdS radius. The Newton constant, $G_3$, and the central charge, $c$, of the dual CFT$_2$ are related by the Brown-Henneaux relation $R_{AdS}/4G_3 = c/6$ \cite{BH86}. 

The solution corresponding to the ground state is global AdS$_3$, given by the metric 
\begin{align}
     ds^2 = -\left(1+\frac{r^2}{R_{AdS}^2}\right) d\tau^2 + \frac{ dr^2}{\left(1+\frac{r^2}{R_{AdS}^2}\right)} + r^2  d\phi^2. 
\end{align}
where $\phi$ is identified with period $2\pi$. To match the convention on the CFT side, we consider that the ADM mass of global AdS$_3$ is $E = -R_{AdS}/8G_3 = -c/12$. Since this is a smooth horizonless geometry, its entropy and free energy are
\begin{equation}
    S_{GAdS_3}\= 0 \,,\qquad F_{GAdS_3} \= -\frac{c}{12}\,.
\end{equation}

In the low-energy range, $-c/12 \leq E \leq 0$, the only solutions are global AdS$_3$ and those with a reparameterization mode on the boundary (called boundary gravitons) on the top of it.\footnote{We could also consider geometries derived from global AdS$_3$ by imposing a conical defect at $r=0$. These solutions have zero entropy and an energy in the range $-c/12 \leq E \leq 0$. However, they involve a point-mass stress-energy tensor and are therefore not typically regarded as pure-gravity solutions. Including these geometries, however, will not affect the main conclusions.} They correspond to the identity primary state and its descendants. 
Consequently, the corresponding dual CFT$_2$ satisfies the sparse-spectrum condition \eqref{eq:sparse_spectrum}. 

At positive energies $E \geq 0$, the static BTZ black hole \cite{Banados:1992wn} provides the relevant solution, with entropy and free energy given by
\begin{equation}
    S_\text{BTZ} = 2\pi \sqrt{\frac{c}{3}E}\,,\qquad F_\text{BTZ} \= -\frac{c}{3} \pi^2 T^2.
\end{equation}
The metric will be presented within its type IIB embedding in a later section.

In the right plot of Fig.\ref{fig:CFTbootstrap_F}, we compare the free energies of global AdS$_3$ and the static BTZ black hole as functions of the temperature. Pure Einstein gravity accurately reflects the universal CFT behavior in canonical ensembles, with a phase transition at $T=(2\pi)^{-1}$: at lower temperatures, global AdS$_3$ dominates, while at higher temperatures, the BTZ black hole dominates.

The right plot in Fig.\ref{fig:CFTbootstrap} shows the entropy as a function of the energy for these configurations. Although $S(E)$ is concave when limited to BTZ black holes, it is not concave across the entire energy range that includes the global AdS$_3$ solution. Thus, the equivalence of statistical ensembles is not expected, implying that the canonical ensemble and the microcanonical ensemble may have different phase structures. 

Pure Einstein gravity does not exhibit exotic behavior in the low- and medium-energy ranges of the microcanonical ensemble, $-c/12 \leq E \leq c/12$, and aligns well with universal CFT predictions. The BTZ black hole entropy reaches the lower bound in the medium-energy range, while global AdS$_3$ is located at the lowest end of the low-energy range.

\section{Localized black holes in Type IIB supergravity}
\label{Sec: Sugra solution}

We now consider the AdS$_3$/CFT$_2$ correspondence in string theory through the ``D1-D5 system" and derive the energy spectrum and possible new saddle points in the low- and medium-energy ranges. We construct supergravity solutions that have AdS$_3 \times$ S$^3 \times$ T$^4$ asymptotics and charges corresponding to $N_1$ D1-branes and $N_5$ D5-branes. The dual description is a two-dimensional CFT with central charge $c=6 N_1 N_5$, known as the D1-D5 CFT (see \cite{Avery:2010qw} for a review).

We parametrize the AdS$_3$ circle using a coordinate $y$ with radius $R_y$, denote the string coupling by $g_s$, the string length by $l_s$, and the T$^4$ volume by $(2\pi)^4 V_4$.

\subsection{The BTZ black hole}
\label{sec:BTZ}

In string frame, the static and spherically symmetric black hole solution with AdS$_3\times$S$^3\times$T$^4$ asymptotics is 
\begin{align}
 ds_{10}^2 \= &\frac{1}{\sqrt{Q_1 Q_5}} \left[-r^2\, dt^2 +(r^2+\ell^2) \, dy^2 \right] + \sqrt{Q_1 Q_5} \left[\frac{ dr^2}{r^2+\ell^2} + d\Omega_3^2 \right]+ \sqrt{\frac{Q_1}{Q_5}}\, ds(T^4)^2  \,,\nn\\
C^{(2)} \= & Q_5 \cos^2 \theta \,d\varphi_2 \wedge  d\varphi_1 -\frac{r^2+\ell^2}{Q_1} \, dt\wedge  dy \,,\qquad  e^\Phi \= \sqrt{\frac{Q_1}{Q_5}}\,,\label{eq:metBTZ}
\end{align}
where $\Phi$ is the dilaton, $C^{(2)}$ is the Ramond-Ramond two-form gauge field, $Q_1$ and $Q_5$ are the D1 and D5 charges respectively, and 
\begin{equation}
\label{Hopf}
d\Omega_3^2=d\theta^2 + \cos^2 \theta \,d\varphi_1^2+ \sin^2 \theta\,d\varphi_2^2
\end{equation}
is the line element of a round S$^3$ in the Hopf coordinates:
\begin{equation}
    0\leq \theta \leq \frac{\pi}{2}\,,\qquad \varphi_1 \= \varphi_1 + 2\pi\,,\qquad \varphi_2 \= \varphi_2 + 2\pi\,. 
\end{equation}
When reduced on S$^3\times$T$^4$ to three dimensions, this solution becomes the static BTZ black hole \cite{Banados:1992wn}. 
At the boundary, this solution asymptotes to AdS$_3\times$S$^3\times$T$^4$:
\begin{equation}
     ds_{10}^2 \sim \sqrt{Q_1 Q_5} \left[ \frac{d\bar{r}^2}{\bar{r}^2} +\bar{r}^2 \left(- d\tau^2 +  d\phi^2 \right)+d\Omega_3^2\right] + \sqrt{\frac{Q_1}{Q_5}}\,ds(T^4)^2\,,
    \label{eq:metAsymp}
\end{equation}
where we have introduced the AdS time, circle and radial coordinates:
\begin{equation}
    \tau \equi \frac{t}{R_y}\,,\qquad \phi \equi \frac{y}{R_y}\,,\qquad \bar{r}^2 \equi \frac{R_y^2}{Q_1 Q_5}\,r^2\,.\label{eq:AsympTime&Y}
\end{equation}
Thus, the AdS radius $R_{AdS}$ is equal to the S$^3$ radius and given by $R_{AdS}=(Q_1 Q_5)^\frac{1}{4}$. The central charge is
\begin{equation}
    c \= 6 N_1 N_5 \= \frac{3\pi\, Q_1 Q_5}{2G_5\, R_y}\,,\label{eq:centralcharge}
\end{equation}
where $N_1$ and $N_5$ are the quantized brane charges:
\begin{equation}
    N_1=\frac{Q_1 V_4}{g_s l_s^6}\,,\qquad N_5=\frac{Q_1}{g_s l_s^2}\,.
    \label{eq:quantizedcharges}
\end{equation}
and $G_5$ is the five-dimensional Newton constant, expressed in terms of the ten-dimensional Newton constant as $G_{10}=8\pi^6 g_s^2 l_s^8$, $G_5= \frac{G_{10}}{(2\pi)^5 V_4 R_y}$.

The black hole has an event horizon of S$^1\times$S$^3\times$T$^4$ topology at $r=0$. The temperature, entropy, and energy of the solution are given by\footnote{The temperature can be derived by regularizing the Euclidean version of the solution at $r=0$, with the periodicity $i\tau = i\tau + T^{-1}$. The energy is derived by expanding the metric near the boundary, which will be explained in detail in Section \ref{sec:ConservedEnergy}.}
\begin{equation}
    T \= \frac{\ell\,R_y}{2\pi \sqrt{Q_1 Q_5}}\,,\qquad S_\text{BTZ} \= \frac{\pi c}{3} \frac{R_y \ell}{\sqrt{Q_1 Q_5}}\,,\qquad E\= \frac{c}{12} \frac{R_y^2\ell^2}{Q_1 Q_5}\,.
    \label{eq:BTZEnergyTemp}
\end{equation}

The free energy of the BTZ black hole is therefore:\footnote{We derive the free energy using the Legendre transformation $F = E - TS$. A more fundamental approach involves computing the free energy directly from the on-shell gravitational path integral. While these methods yield the same result for Einstein gravity coupled to matter fields, like type IIB supergravity, they generally produce different results in the presence of higher-derivative corrections to gravity \cite{Wald:1993nt,Jacobson:1993vj,Iyer:1995kg}. In this paper, we adopt the first approach as it is significantly simpler to compute.}
\begin{equation}
    F_\text{BTZ}\equi  E-T S_\text{BTZ} \=-\frac{\pi}{8}\,\frac{R_y \ell^2}{G_5}\quad \Rightarrow\quad F_\text{BTZ} \= -\frac{c}{3}\, \pi^2T^2\,.
    \label{eq:FreeEnergyBTZ}
\end{equation}

The parameter $\ell$ corresponds to the non-extremality parameter of the black hole and to the size of the horizon. For $\ell=0$, we retrieve the extremal massless BTZ solution with zero temperature, entropy, and energy, and the metric turns into Poincar\'e AdS$_3$.

\subsection{The localized black holes}
\label{sec:LocBHGen}

In \cite{Bah:2022pdn}, a novel integrable structure in type II supergravity was developed, based on the generalization of the electrostatic Ernst formalism in supergravity \cite{Heidmann:2021cms,Bah:2021owp,Bah:2021rki,Heidmann:2022zyd,Bah:2022yji,Bah:2023ows,Heidmann:2023kry}. This has enabled the construction of novel non-BPS static and axially symmetric geometries in AdS$_3$, including smooth horizonless bubbling geometries and regular bound states of black holes. In this paper, we focus exclusively on the black geometries.

The new solutions share the same AdS$_3\times$S$^3\times$T$^4$ asymptotic structure as the BTZ black hole, but depend non-trivially not only on  $r$, the AdS radial coordinate, but also on $\theta$, the angular position on the S$^3$. 
For the BTZ black hole, only the time coordinate degenerates, and this happens at the horizon ($r=0$), regardless of the value of $\theta$. Thus, the BTZ horizon is uniform over the S$^3$. In contrast, for the new solutions, various coordinates degenerate as one moves on the S$^3$ along $\theta$: either the time direction degenerates, indicating the presence of an event horizon, or the AdS$_3$ $y$-circle degenerates smoothly, as it happens at the center of global AdS$_3$.

\subsubsection{The solutions}
\label{eq:LocBHSol}

The localized multi-black hole solutions are characterized by $n$ colinear line sources that we can take to lay at the origin of the radial coordinate, $r=0$. At these sources the time coordinate degenerates, and hence the solution describes a bound state of black holes that localize in various regions of the S$^3$. The black holes are separated by regions where the AdS$_3$ circle degenerates smoothly.\footnote{In \cite{Bah:2022pdn}, more generic geometries have been constructed where the spacelike coordinates that degenerate between black holes can be a T$^4$ coordinate or an S$^3$ angle} We introduce spherical coordinates centered around each source, $(r_i, \theta_i)$:
\begin{align}
r_i^2 \equi & \rho_i +\rho_{i-1} - \frac{\ell_i^2}{2}\,,\qquad \ell_i^2 \cos^2\theta_i \equi \rho_{i-1} - \rho_{i} + \frac{\ell_i^2}{2}\,,\nn\\ 
\rho_i \equi & \frac{1}{4} \, \sqrt{\left((2r^2+\ell^2) \,\cos 2\theta+\ell^2 -2\sum_{j=1}^{i}\ell_j^2 \right)^2+4r^2(r^2+\ell^2)\sin^22\theta}\,, \label{eq:ri&thetaidef}
\end{align}
where $\ell_i$ corresponds to the size of the $i^\text{th}$ source, while $\ell$ is the size of the entire solution:
\begin{equation}
    \ell^2 = \sum_{i=1}^n \ell_i^2\,.\label{eq:SolutionLength}
\end{equation}
One can check that at $r=0$, a single $r_i$ vanishes while all other $r_j$ remain finite, depending on the values of $\theta$:
\begin{equation}
    r=0\,,\quad \frac{1}{\ell^2}\sum_{j=1}^{i-1}\ell_j^2 \,<\,  \cos^2\theta \,<\, \frac{1}{\ell^2}\sum_{j=1}^{i}\ell_j^2 \qquad \Leftrightarrow \qquad r_i=0\,,\quad r_j >0 \,,\quad j\neq i\,.
    \label{eq:AngleRanges}
\end{equation}
Each of these loci corresponds to either a black hole horizon or a smooth degeneracy of the AdS$_3$ circle. For this purpose, we define the sets of labels, $U_t$ and $U_y$, so that:
\begin{equation}
\begin{split}
    &i \in U_t \quad \Rightarrow \quad r_i=0 \text{ corresponds to a black hole horizon,} \\
    &i \in U_y \quad \Rightarrow \quad r_i=0 \text{ corresponds to a smooth degeneracy.} \\
\end{split}
\end{equation}

Finally, the type IIB metric is given by
\begin{align}
 ds_{10}^2 = &  \frac{1}{\sqrt{Q_1 Q_5}} \left[-r^2 \,\cK_y(r,\theta) \,dt^2 +\frac{r^2+\ell^2}{\cK_y(r,\theta)} \,dy^2 \right] + \sqrt{\frac{Q_1}{Q_5}}\, ds(T^4)^2\nn \\
&+ \sqrt{Q_1 Q_5} \,\Biggl[ G(r,\theta)\,\left(\frac{ dr^2}{r^2+\ell^2}+d\theta^2 \right) + \cos^2 \theta \,d\varphi_1^2+ \sin^2 \theta\,d\varphi_2^2 \Biggr]\,,\label{eq:metAdS3+BTZs} 
\end{align}
while the RR field, $C^{(2)}$, and the dilaton, $\Phi$, are identical to the BTZ solution \eqref{eq:metBTZ}. The warp factor $\cK_y$ indicates the loci where the $y$-circle degenerates, and the warp factor $G$ can be interpreted as capturing the one-to-one interactions between the sources:
\begin{equation}
\cK_y \equiv \prod_{i\in U_y}\left( 1+\frac{\ell_i^2}{r_i^2}\right), \quad G \equiv \prod\limits_{\substack{i\in U_y\\j\in U_t}}\frac{\left( \left(r_i^2+\ell_i^2 \right) \cos^2\theta_i + r_j^2 \sin^2\theta_j \right)\left(r_i^2 \cos^2\theta_i +  \left(r_j^2+\ell_j^2\right) \sin^2\theta_j \right)}{\left( \left(r_i^2+\ell_i^2 \right) \cos^2\theta_i +  \left(r_j^2+\ell_j^2\right) \sin^2\theta_j \right)\left(r_i^2 \cos^2\theta_i +  r_j^2 \sin^2\theta_j \right)}.\label{eq:FieldGeneric}
\end{equation}

\begin{figure}
    \centering
    \includegraphics[width=0.9 \textwidth]{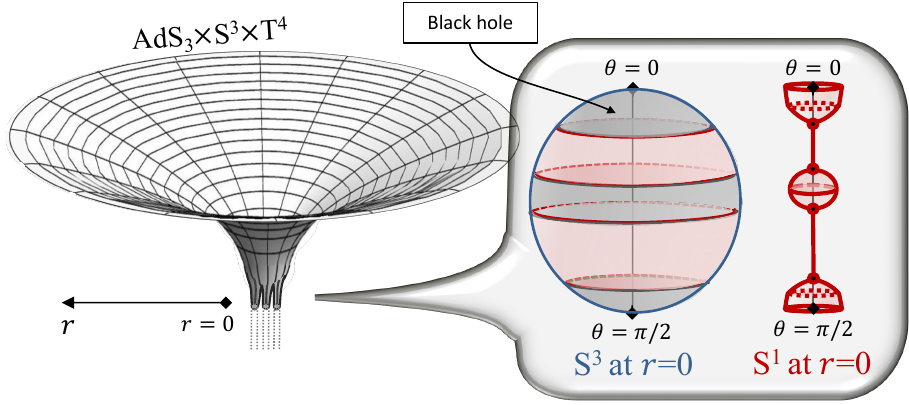}
    \caption{Schematic description of our black-hole bound-state geometries. The S$^3$ at $r=0$ is covered in red and gray regions. In the red regions the $y$-circle degenerates, and the spacetime caps off smoothly. In the gray regions the $t$-circle degenerates so we have an event horizon. In this example, we have three disconnected gray regions, and therefore three disconnected horizons.}
    \label{fig:LocalizedBH}
\end{figure}

The solutions are asymptotic to AdS$_3 \times$ S$^3 \times$ T$^4$. Moreover, one can verify that the spacetime is regular for $r > 0$, and the $(\theta, \varphi_1, \varphi_2)$ coordinates define a stretched S$^3$ since $G(r, 0) = G(r, \pi/2) = 1$. At $r = 0$, either the time fiber or the $y$ fiber degenerates depending on the position on the S$^3$ and the specific $r_i$ that vanishes there. The spacetime structure of the solutions is depicted in Fig.\ref{fig:LocalizedBH}.

This class of solutions includes the BTZ black hole ($n = 1$ with $U_t = \{1\}$ and $U_y = \emptyset$), as well as global AdS$_3$ ($n = 1$ with $U_t = \emptyset$ and $U_y = \{1\}$) as special limits.

\subsubsection{Conserved charges, energy and free energy}
\label{sec:ConservedEnergy}

At large distances, $r \to \infty$, $\cK_y$ and $G$ tend to 1, so the solutions asymptotically approach AdS$_3 \times$ S$^3 \times$ T$^4$, as one can see from  \eqref{eq:metAsymp}. The central charge and the quantized D1-brane and D5-brane charges are given in \eqref{eq:centralcharge} and \eqref{eq:quantizedcharges}

To calculate the energy of CFT states dual to our solutions from their bulk asymptotics we apply the method outlined in \cite{Kanitscheider:2007wq,Kraus:2006wn,Ganchev:2021ewa}, summarized here in the static limit. We first move to the Fefferman-Graham gauge, where the metric takes the asymptotic form:
\begin{equation}
     ds^2=\frac{\sqrt{Q_1Q_5}}{u^2}\left(du^2- d\tau^2+ d\phi^2+u^2\,g_{\mu\nu}^{(2)}dx^{\mu}dx^{\nu}\right)+ \sqrt{Q_1Q_5} d\Omega_3^2 +\sqrt{\frac{Q_1}{Q_5}} \, ds(T^4)^2\,, 
\end{equation}
where $u=\frac{\sqrt{Q_1 Q_5}}{r R_y} \to 0$, and $x^\mu=(\tau,\phi)$ are defined in \eqref{eq:AsympTime&Y}. The left and right conformal dimensions are given by
\begin{equation}
    h-\frac{c}{24} \=\frac{c}{24}\left(g_{\phi\phi}^{(2)}+g_{\tau\tau}^{(2)}+2g_{\tau\phi}^{(2)}\right)\,\qquad \bar{h}-\frac{c}{24} \=\frac{c}{24}\left(g_{\phi\phi}^{(2)}+g_{\tau\tau}^{(2)}-2g_{\tau\phi}^{(2)}\right)\,,
\end{equation}
and the energy is given by:\footnote{Note that our solutions have no momentum charge $N_p=h-\bar{h}$, since $g_{\tau \phi}=0$.}
\begin{equation}
    E \= h+\bar{h}-\frac{c}{12}\,,
\end{equation}
where $-\frac{c}{12}$ is the energy of global AdS$_3$, the solution dual to the NS-NS ground state of the D1-D5 CFT.

After expanding the metric \eqref{eq:metAdS3+BTZs}, we find that the energy of our solutions is 
\begin{equation}
    \label{eq:Energyboundstate}
    E\=\frac{c\,R_y^2}{12\, Q_1 Q_5}\,\left(\sum_{i\in U_t}\ell_i^2-\sum_{i\in U_y} \ell_i^2\right)\,.
\end{equation}
As expected, we recover the BTZ energy \eqref{eq:BTZEnergyTemp} when $U_t = \{1\}$ and $U_y = \emptyset$, and the global AdS$_3$ energy when $U_t = \emptyset$ and $U_y = \{1\}$.\footnote{For global AdS$_3$, the regularity condition where the $y$ circle degenerates imposes $\ell^2=\frac{Q_1 Q_5}{R_y^2}$.}
For generic configurations, each black hole contributes positively to the energy, while each $y$-circle degeneracy gives a negative contribution. 

Moreover, anticipating what we will derive next, the entropy of the solutions in which all the black holes are in thermal equilibrium at temperature $T$, is given by:
\begin{equation}
    S \= \frac{\pi\,R_y}{4 T G_5}\,\sum_{i\in U_t}\ell_i^2\,.
    \label{eq:EntropyBoundState}
\end{equation}
Thus, the free energy, expressed in terms of $\ell$ \eqref{eq:SolutionLength}, matches the free energy of the BTZ black hole \eqref{eq:FreeEnergyBTZ}:
\begin{equation}
    F \= E-T S\= -\frac{\pi R_y\,\ell^2}{8 G_5}\,.\label{eq:FreeEnergyBoundState}
\end{equation}
Note, however, that its dependence as a function of the temperature may differ significantly from that of the BTZ free energy, as we will analyze in detail in Section \ref{sec:PhaseDiag}.

\subsubsection{Internal structure}

At $r=0$, either the time fiber or the $y$ fiber degenerates, depending on the position along the $S^3$ and on which of the $r_i$ vanishes.  To ensure that the black holes are in thermal equilibrium and that the degeneracy of the $y$-circle corresponds to a smooth origin in $\mathbb{R}^2$, it is necessary to impose $n$ algebraic constraints that fix all lengths $\ell_i^2$ in terms of the temperature, $T$, and the S$^1$ radius, $R_y$. For the derivation of these constraints, we refer the interested reader to \cite{Bah:2022pdn}. The final result is:
\begin{align}
R_y\=\frac{\sqrt{Q_1 Q_5}\,\ell_i\,d_i}{\ell^2}  \,,\quad \text{if } i\in U_y\,, \qquad T \= \frac{R_y \ell^2}{2\pi\sqrt{Q_1 Q_5} \,\ell_i \,d_i} \,,\quad\text{if } i\in U_t,\label{eq:RegConstraintsGen}
\end{align}
where we have defined the aspect ratios:
\begin{equation}
d_i \equi \prod_{p=1}^{i-1} \prod_{q=i+1}^n \left[\frac{1+ \frac{\ell_q^2}{\sum_{k=p}^{q-1} \ell_k^2}}{1+ \frac{\ell_q^2}{\sum_{k=p+1}^{q-1} \ell_k^2}} \right]^{\frac{\alpha_{pq}}{2}} \,\prod_{p=1}^{i-1}\left(1+ \frac{\ell_p^2}{\sum_{k=p+1}^{i} \ell_k^2} \right)^{\frac{\alpha_{ip}}{2}} \, \prod_{p=i+1}^{n}\left(1+ \frac{\ell_p^2}{\sum_{k=i}^{p-1} \ell_k^2} \right)^{\frac{\alpha_{ip}}{2}}\,, \label{eq:DefdiAspect}
\end{equation}
and $\alpha_{ij}=1$ if $i,j\in U_w$ or $0$ if $i\in U_w$ and $j\in U_{w'}$ with $w\neq w'$. Unfortunately, such constraints are not analytically solvable except for solutions with a small number of sources, $n=2$ and $n=3$. Additionally, each source carries D1 and D5 charges given in terms of the total charges by \cite{Bah:2022pdn}, 
\begin{equation}
    q_1^{(i)} \= \frac{\ell_i^2}{\ell^2}\, Q_1\,,\qquad q_5^{(i)} \= \frac{\ell_i^2}{\ell^2}\, Q_1\,.
\end{equation}

Finally, the Bekenstein-Hawking entropy, $S_i$, of each black hole in the bound state is
\begin{equation}
    S_i \= \frac{\pi^2 \sqrt{Q_1 Q_5}\,\ell_i^2 d_i}{2G_5\,\ell^2}\= \frac{\pi \ell_i^2 R_y}{4 G_5 T}\,,
\end{equation}
which leads to the total bound state entropy $S=\sum_{i\in U_t} S_i$, as specified in equation \eqref{eq:EntropyBoundState}.

\subsection{Simple examples}
\begin{figure}
    \centering
    \includegraphics[width=0.9 \textwidth]{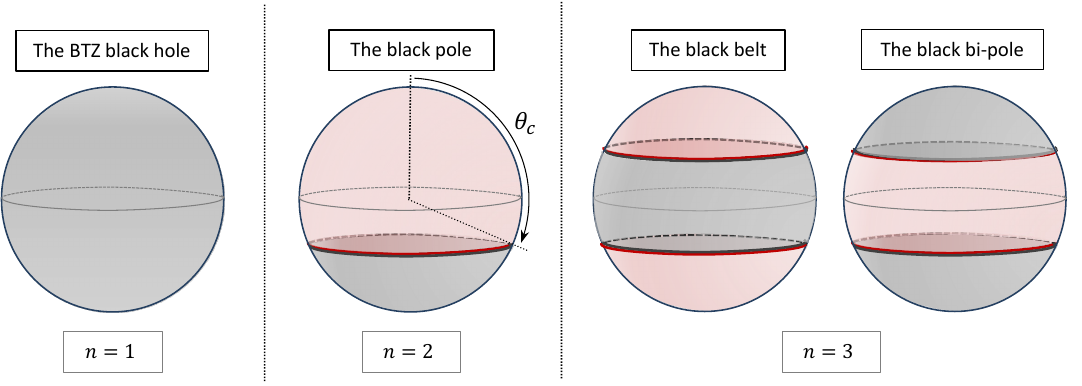}
    \caption{Representation of the S$^3$ regions for the simplest localized black hole geometries. The BTZ black hole, where the horizon is uniform along the S$^3$. The black pole, where the black hole is centered around one of the poles of the sphere (in Hopf coordinates \eqref{Hopf}). The black belt and the black bi-pole are $\mathbb{Z}_2$ symmetric solutions with either one black belt centered on the equator or two symmetric black poles at the poles of the S$^3$.}
    \label{fig:BHSol}
\end{figure}

In this section, we construct several (relatively simple) examples of solutions belonging to the general class of localized black holes introduced in the previous section. These examples, depicted in Fig.\ref{fig:BHSol}, involve solutions with a small number of sources. In addition to the BTZ black hole obtained with $n=1$ and reviewed in Section \ref{sec:BTZ}, we consider:
\begin{itemize}
    \item \underline{$n=2$:} The black pole.
    
    For two sources, the only solution corresponds to a black hole centered around a pole of the S$^3$, while the other region of the S$^3$ corresponds to the smooth degeneracy of the AdS$_3$ angle.
    \item \underline{$n=3$:} The black belt and the black bi-pole.

    For three sources, there are two types of solutions: the black belt, with a grey region corresponding to an event horizon centered on the equator of the S$^3$, and the black bi-pole, with two identical black holes at the poles of the S$^3$.
\end{itemize}
The metric and fields for each solution can be derived using the generic form \eqref{eq:metAdS3+BTZs}. In this section, we will specify the conserved and thermodynamic quantities by solving the regularity constraints \eqref{eq:RegConstraintsGen}.

\subsubsection{The black pole}

The black pole ($n=2\,, U_t \= \{1\}\,, U_y \= \{2\}\,$)
is the simplest solution within the class of geometries introduced in Section \ref{sec:LocBHGen} that breaks the $\rm S^3$ isometry. 

The regularity constraints \eqref{eq:RegConstraintsGen} are invertible and yield:
\begin{equation}
    \ell^2 \= \frac{Q_1 Q_5}{R_y^2}\,\frac{4\pi^2 T^2}{1 +4\pi^2 T^2}\,,\qquad \ell_1^2 \= \frac{\ell^2}{1 +4\pi^2 T^2}\,,\qquad \ell_2^2 \= \frac{4\pi^2 T^2\,\ell^2}{1 +4\pi^2 T^2}\,.
    \label{eq:LengthRodBP}
\end{equation}
Thus, there exists a single solution for a given temperature, AdS radius, and S$^1$ radius.

At high temperatures ($T\gg 1$), $\ell_1^2\to 0$ while $\ell_2^2\sim \ell^2 \to \frac{Q_1 Q_5}{R_y^2}$. Consequently, the black hole shrinks, and the region where the $y$-circle degenerates occupies the entire S$^3$. The solution converges towards a global AdS$_3$ spacetime. At low temperatures ($T\ll 1$), both lengths shrink, but $\ell_2^2$ vanishes more rapidly, leading the solution to converge towards the zero-energy singular BTZ black hole. For intermediate temperatures, the solution represents a novel black geometry corresponding to a black pole on the S$^3$, as illustrated in Fig.\ref{fig:BHSol}, with the region of the S$^3$ ``covered'' by the black hole of size $\ell_1^2/\ell^2 = (1 +4\pi^2 T^2)^{-1}$.

Note that in our type IIB solutions the black hole localized on the S$^3$ has the same horizon topology as the BTZ black hole: S$^3\times$S$^1\times$T$^4$. The difference is that coordinates that parameterize the BTZ horizon S$^3$ are $(\theta,\varphi_1,\varphi_2)$ while the coordinates parameterizing the black-pole horizon S$^3$ are $(\theta,\varphi_1,y)$. 

We can express the energy \eqref{eq:Energyboundstate}, entropy \eqref{eq:EntropyBoundState}, and free energy \eqref{eq:FreeEnergyBoundState} of the solution as functions of the temperature:
\begin{equation}
    E \= \frac{c}{12}\,\frac{4\pi^2 T^2 \left(1-4\pi^2T^2 \right)}{\left(1+4\pi^2T^2 \right)^2}\,, \qquad S\= \frac{c}{6} \,\frac{4\pi^2 T}{\left(1+4\pi^2T^2 \right)^2}\,,\qquad F\= - \frac{c}{12}\,\frac{4\pi^2 T^2}{1+4\pi^2T^2}\,.
    \label{eq:LocBHEnEntFreeEn}
\end{equation}

The expression for the energy indicates that the solution only exists in the energy range:
\begin{equation}
     -\frac{c}{12}= E_{G\text{AdS}_3}\,<\, E \,\leq\, E_\text{max} \= \frac{c}{96}\,,
    \label{eq:EnergyRangeLocBH}
\end{equation}
where the upper bound is achieved for $T=T_c\equiv (2\sqrt{3}\pi)^{-1}$, and $E_{G\text{AdS}_3}=-\frac{c}{12}$ is the energy of global AdS$_3$.

In the language introduced in Section \ref{sec:CFT}, the black pole solution can only be found in the 
low-energy range ($-\frac{c}{12}\leq E\leq0$) and the medium-energy range ($0\leq E\leq\frac{c}{12}$).

The solutions can also be given at fixed energy instead of fixed temperature which will be particularly useful when we discuss the microcanonical ensemble in Section \ref{sec:PhaseDiag}. Within the energy range \eqref{eq:EnergyRangeLocBH}, we find two branches of solutions at positive energy, with temperatures $T_\pm$, and a single solution at negative energy, with temperature $T_-$:
\begin{equation}
    4\pi^2 T_\pm^2 \= \frac{4}{3\pm \sqrt{1-\frac{E}{E_\text{max}}}}-1. 
    \label{eq:TwoBranchesLocBH}
\end{equation}
The entropy of both black-pole branches as a function of energy is
\begin{equation}
    S^\pm \= \frac{\pi c}{48} \left(3 \pm \sqrt{1-\frac{E}{E_\text{max}}} \right)^2 \sqrt{ \frac{4}{3\pm \sqrt{1-\frac{E}{E_\text{max}}}}-1},
    \label{eq:BlackPoleEntropIn}
\end{equation}
and one can check that the minus branch corresponds to a solution with nonzero entropy that exists in the low-energy range, $-\frac{c}{12}\leq E\leq0$, where the BTZ black hole does not exist.

\subsubsection{The black belt}

The black belt solution is given by the following configuration: $n=3$, $U_t = \{2\},$ and $U_y \= \{1,3\}$.
As illustrated in Fig.\ref{fig:BHSol}, the gray region corresponding to the black-hole horizon is enclosed by two regions where the $y$-circle degenerates. This black-belt horizon topology is S$^2\times$S$^1\times$S$^1\times$T$^4$. This differs from the BTZ and black-pole horizon topology, and one can think about this belt as a black string wrapping the equator of the $\rm S^3$.

The regularity constraints \eqref{eq:RegConstraintsGen} yield:
\begin{equation}
    \ell^2 = \frac{Q_1 Q_5}{R_y^2} \frac{\sqrt{1+16\pi^2T^2}-1}{\sqrt{1+16\pi^2T^2}+1}\,,\quad \ell_1^2=\ell_3^2= \frac{\ell^2}{2}\left(1-\frac{1}{\sqrt{1+16\pi^2T^2}} \right),\quad \ell_2^2 =\frac{\ell^2}{\sqrt{1+16\pi^2T^2}}.
\end{equation}
Thus, the solution has a similar behavior to the black pole geometry at both high and low temperatures. It converges towards global AdS$_3$ at high temperatures and approaches the zero-energy singular BTZ black hole at low temperatures.

The energy \eqref{eq:Energyboundstate}, entropy \eqref{eq:EntropyBoundState}, and free energy \eqref{eq:FreeEnergyBoundState} of the solution are, as functions of the temperature,
\begin{equation}
\begin{split}
    E &\= \frac{c}{12}\,\frac{\sqrt{1+16\pi^2T^2}-1}{\sqrt{1+16\pi^2T^2}+1} \left( \frac{2}{\sqrt{1+16\pi^2T^2}}-1\right)\,, \\
    S&\= \frac{c\,\left(\sqrt{1+16\pi^2T^2}-1 \right)^2}{96\pi^2 T^3\sqrt{1+16\pi^2T^2}} \,,\qquad F\= - \frac{c}{12}\,\frac{\sqrt{1+16\pi^2T^2}-1}{\sqrt{1+16\pi^2T^2}+1}\,.
    \label{eq:BlackBeltEnEntFreeEn}
\end{split}
\end{equation}
Like the black pole, the black belt exists in a limited energy range:
\begin{equation}
  -\frac{c}{12} =  E_{G\text{AdS}_3} \,<\, E\,\leq\, \frac{c}{12}\left(7-4\sqrt{3} \right).
\end{equation}
Notably, this energy range is smaller than that of the black pole \eqref{eq:EnergyRangeLocBH}, with the upper bound being approximately $0.6 E_\text{max}$.

Finally, one can express the solution in terms of energy $E$ instead of temperature. As with the black pole, there are two solutions corresponding to two different temperatures, $T_\pm$, for positive energy, while only the minus branch exists for negative energy.

\subsubsection{The black bi-pole}

The black bi-pole solution, depicted in Fig.\ref{fig:BHSol}, is characterized by: $n=3$, $U_t = \{1,3\}$ and $U_y = \{2\}$.
This solution has two horizons localized at both poles of the S$^3$ (in Hopf coordinates \eqref{Hopf}). These are black poles with an S$^3\times$S$^1\times$T$^4$ horizon topology, where the two three-spheres are spanned by $(\theta,\varphi_1,y)$ and $(\theta,\varphi_2,y)$ respectively.

The regularity constraints \eqref{eq:RegConstraintsGen} yield:
\begin{equation}
    \ell^2 = \frac{4\pi^2 Q_1 Q_5\,T^2\,\left(\sqrt{1+\pi^2T^2}-\pi T \right)^2}{R_y^2}\,,\quad \ell_1^2=\ell_3^2= \frac{\ell^2}{2}\left(1-\frac{\pi T}{\sqrt{1+\pi^2T^2}} \right),\quad \ell_2^2 =\frac{\pi \ell^2\,T}{\sqrt{1+\pi^2T^2}}.
\end{equation}
As before, the solution approaches global AdS$_3$ at high temperatures and the zero-energy singular BTZ geometry at low temperatures. The energy \eqref{eq:Energyboundstate}, entropy \eqref{eq:EntropyBoundState}, and free energy \eqref{eq:FreeEnergyBoundState} of the solution are, as functions of the temperature,
\begin{equation}
\begin{split}
    E &\= \frac{ c \,\pi^2T^2\, \left(\sqrt{1+\pi^2T^2}-\pi T \right)^2}{3}\, \left( 1-\frac{2\pi T}{\sqrt{1+\pi^2T^2}}\right)\,, \\
    S&\= \frac{2c\,\pi^2T^2\,\left(\sqrt{1+\pi^2T^2}-\pi T \right)^3}{3\sqrt{1+\pi^2T^2}} \,,\qquad F\= - \frac{c}{3}\,\pi^2 T^2\left(\sqrt{1+\pi^2T^2}-\pi T \right)^2\,.
    \label{eq:DoubBHEnEntFreeEn}
\end{split}
\end{equation}
The black bi-pole solution exists within the energy range:
\begin{equation}
  -\frac{c}{12}=    E_{G\text{AdS}_3} \,<\, E\,\leq\, \frac{c}{18}\left(47-13\sqrt{13} \right).
\end{equation}
Since the properties of the solution are similar to the previous solutions, further details can be found in previous sections.

\subsubsection{More fragmented solutions and static black Saturns}
\label{sec:BlackSaturns}

The class of localized black holes constructed in section \ref{sec:LocBHGen} encompasses more sophisticated solutions with additional sources at the $r=0$ locus. This results in geometries with multiple horizons on the S$^3$, separated by regions where the $y$-circle smoothly degenerates. These configurations can be viewed as AdS$_3$ analogs of static black Saturns \cite{Bena:2004de,Elvang:2007rd}, where the $r=0$ locus forms a bound state of one or two black holes with multiple black strings in their vicinity (as depicted in Fig.\ref{fig:LocalizedBH}).

Unfortunately, for these geometries, one can only solve the regularity constraints \eqref{eq:RegConstraintsGen} numerically. However, we observed that these solutions exhibit thermodynamic properties similar to those of the simpler geometries discussed above. Specifically, there is a unique solution for a given temperature, AdS radius, and S$^1$ radius, and these solutions converge towards global AdS$_3$ at high temperatures and towards the zero-energy singular BTZ black hole at low temperatures. Additionally, their energies are constrained within a similar range as \eqref{eq:EnergyRangeLocBH}, though the upper bound is different, and there are two possible solutions at fixed positive energy. 

\section{Thermodynamics at low energy}
\label{sec:ThermoLowEn}

The supergravity solutions discussed in the previous section introduce new thermodynamic phases and saddles in the low-energy range $-\frac{c}{12} \leq E \leq 0$ between global AdS$_3$ and the zero-energy singular BTZ energies. As reviewed in Section \ref{sec:CFT}, this range is within the canonical shadow where little is known from a CFT perspective, except that the spectrum density should be sparse enough to satisfy \eqref{eq:sparse_spectrum}.

Our ten-dimensional solutions demonstrate that there exist a large number of physical states in the canonical shadow, populating the low-energy spectrum. While pure Einstein gravity in three dimensions sees a large gap between $E = -c/12$ and $E = 0$, with a significant transition at $E = 0$ marked by the emergence of the BTZ solution, nothing of that kind occurs in String Theory.
Instead, the entropy remains $\cO(c)$ from $E>-c/12$ up to $E = 0$, with no indication of a phase transition at $E=0$.
Pure Einstein gravity (reviewed in Section \ref{sec:EinsteinGrav}) fails to capture these states, because they break the S$^3$ isometry. Indeed, from a purely three-dimensional perspective, our solutions have infinite towers of excited KK modes.

\begin{figure}[t]
    \centering
    \includegraphics[width=16cm]{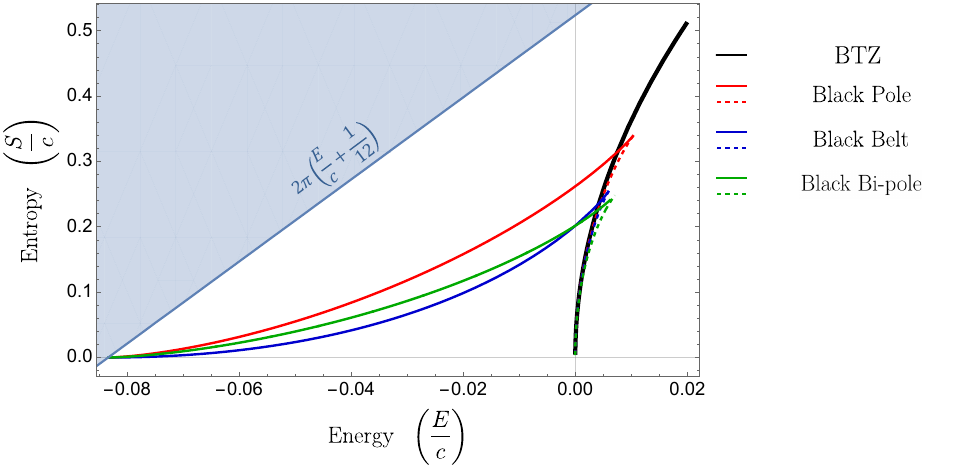}
    \caption{Entropy as a function of energy for several families of localized black hole geometries. For each family, the dotted line corresponds to the plus branch that only exists at $E>0$ (like the BTZ black hole), while the solid line corresponds to the minus branch which also exists in the low-energy range, $E<0$. The region outside the spareness condition \eqref{eq:sparse_spectrum} is shown in blue and its boundary describes the symmetric orbifold D1-D5 CFT, which saturates this condition \cite{Hartman:2014oaa}.}
    \label{fig:NegEn}
\end{figure}

The low-energy spectrum of states in AdS$_3\times$S$^3\times$T$^4$ is therefore described by localized black hole geometries whose entropy scales with the central charge, revealing a vast degeneracy of states. In Fig.\ref{fig:NegEn}, we plot the entropy as a function of the energy for the black pole \eqref{eq:BlackPoleEntropIn}, black belt, and black bi-pole, focusing on this low-energy regime.

As previously discussed, only the minus branches of the black geometries exist at low energies. All of these branches converge toward global AdS$_3$ at $E = -c/12$, as the localized horizons in the geometries shrink to zero size. 

Despite the large number of states, these entropies still satisfy the sparseness condition \eqref{eq:sparse_spectrum} imposed on the CFT side. Thus, while the new ten-dimensional black-hole geometries reveal a substantial number of states in the low-energy range, they remain consistent with a sparse spectrum, as assumed in the large-$c$ modular bootstrap.

Recall that this bound is saturated by symmetric orbifold CFTs \cite{Hartman:2014oaa}, including the weak-coupling limit of the D1-D5 CFT. In 3D pure Einstein gravity, all these low-energy CFT states are not captured in the strong-coupling limit. In contrast, our localized black hole geometries capture a significant fraction of the entropy of these weak-coupling states indicating that a significant number of these states are not lifted at strong coupling.

More precisely, in the low-energy range ($-c/12 < E \leq 0$), the maximum entropy is always carried by the black pole. 
At $E = 0$, its entropy is exactly \emph{half} of the entropy in the weak-coupling limit of the D1-D5 CFT, $S(E=0)=\frac{\pi c}{12} \= \frac{1}{2} S_{\rm CFT}^{\rm (weak)}$. 

At energies slightly above $-c/12$, the S$^3$ region corresponding to the black-pole horizon (shown in Fig.\ref{fig:LocalizedBH}) is very small and the solution looks like global AdS$_3\times$S$^3$ with a tiny black hole near the one of the $\rm S^3$ poles (in Hopf coordinates). As one increases the energy this region grows until it covers half of the S$^3$ at $E = 0$.\footnote{To determine the extent of the S$^3$ region corresponding to the horizon, we apply \eqref{eq:AngleRanges} to compute the angular interval corresponding to either a horizon or a degeneracy of the $y$-circle. This calculation uses the $\ell_i$ values at fixed energy, given in \eqref{eq:LengthRodBP} and \eqref{eq:TwoBranchesLocBH}.}

We would like to emphasize that our construction does not determine whether the remaining states of the free orbifold D1-D5 CFT are lifted. Indeed, there could exist other (potentially more dominant) localized black-hole geometries, which are not included in the ansatz \cite{Heidmann:2021cms,Bah:2022pdn} used to generate our solutions. One such possible geometry is a localized small $\rm AdS_3\times S^3$ black hole that has $\rm S^4$ horizon topology, constructed numerically in \cite{Dias-Santos-toappear}.

\section{Thermodynamics at medium energy}
\label{sec:PhaseDiag}

In the medium-energy range, $0<E<\frac{c}{12}$, the localized black hole geometries can coexist with the BTZ black hole. Thus, we can analyze these solutions in the microcanonical and canonical ensembles and reveal the potential phase transitions. 

We will show that the BTZ black hole does not dominate the microcanonical ensemble below a critical energy $E_c$ and should exhibit a Gregory-Laflamme-type instability in favor of the localized black hole geometries. Among our solutions, the most microcanonically dominant phase is the minus branch of the black pole, which dominates also in the low-energy range. This is fully consistent with the general expectations outlined in \cite{Banks:1998dd}.

However, we will show that the canonical ensemble is blind to this phase transition as expected from the universal CFT arguments reviewed in Section \ref{sec:CFT}: the only phases visible in the canonical ensemble are the BTZ black hole and thermal AdS.

\subsection{Microcanonical ensemble}
\label{sec:MicroCano}

In the microcanonical ensemble, the energy, $E$, the charges, $(Q_1,Q_5)$, and the sizes of the extra dimensions, $(R_y,V_4)$, are held fixed. The most dominant phase is the one that has the maximum entropy, $S$. 

As demonstrated in Section \ref{Sec: Sugra solution}, only the BTZ black hole exists for $E> E_\text{max}=\frac{c}{96}$ \eqref{eq:EnergyRangeLocBH}, and its entropy, as the function of energy, is \eqref{eq:BTZEnergyTemp}:
\begin{equation}
    S_\text{BTZ} \= \frac{2\pi \sqrt{c}}{\sqrt{3}}\,\sqrt{E}\,.
    \label{eq:BTZEntropvsEnergy}
\end{equation}

For $E\leq E_\text{max}$, the localized black hole geometries start to exist, starting with the black pole, then the black bi-pole, the black belt, and all other fragmented black geometries. For each solution, one can calculate their entropy as a function of the energy and determine the most dominant phase in the microcanonical ensemble. \\

We first consider the black pole. At $E\leq E_\text{max}$, there are two different solutions, with entropies \eqref{eq:BlackPoleEntropIn}. The plus branch has always less entropy than the BTZ black hole, while the minus branch becomes more entropic when the energy is below a critical value:
\begin{equation}
 E \,\leq\, E_c \equi \frac{c}{24}\left(5\sqrt{5}-11\right) \sim 0.0075 c\quad \Longrightarrow \quad S_{\text{BP}}^- \,\geq\, S_\text{BTZ}.
\end{equation}
Thus, the BTZ black hole is not the most dominant phase in the microcanonical ensemble for $E\leq E_c$.
\\

\begin{figure}[t]
    \centering
    \includegraphics{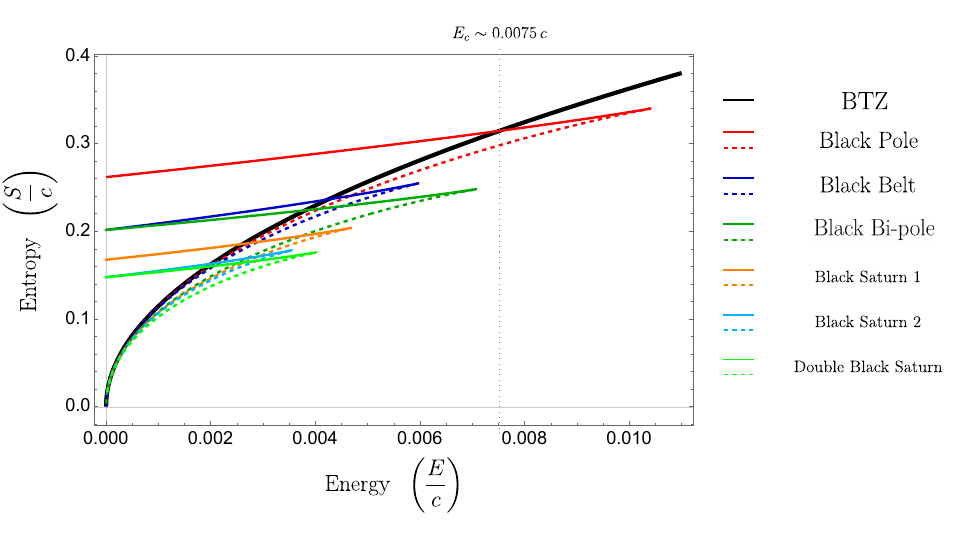}
    \caption{Entropy as a function of the energy for the BTZ black hole and several localized black hole geometries in the medium energy range. For each family, the dotted line corresponds to the plus branch while the solid line corresponds to the minus branch and becomes more entropic than BTZ below a certain energy. The minus branch of the black pole is the dominant phase in the microcanonical ensemble for $E\leq E_c \sim 0.0075\,c$.}
    \label{fig:microcanonical}
\end{figure}

A similar analysis can be performed for each family: the black belt, the black bi-pole, etc. In Fig.\ref{fig:microcanonical}, we have plotted the entropy as a function of the energy for both branches of several families of solutions. We have included some of the more fragmented geometries discussed in Section \ref{sec:BlackSaturns}, including the black Saturn geometries that have multiple horizons along the S$^3$.

The plot shows that each family has the same behavior as the black pole but their overall entropy is smaller as the localized black hole geometries get more and more fragmented. Therefore, the most dominant thermodynamic phase in the microcanonical ensemble is the black pole when the energy is below the critical value $E \leq E_c$ and remains the BTZ black hole for higher energies.  

 Note that the sparseness condition \eqref{eq:medium_up} is satisfied by the most dominant geometries in the medium range: the minus branch of the black pole at $0<E \leq E_c$ and the BTZ black hole at $E_c < E \leq c/12$. Hence our results are consistent with the CFT expectations.
 
The localized black hole geometries provide the direct realization of the expectations put forth in \cite{Banks:1998dd}. A phase transition occurs between the BTZ black hole and the black pole at $E\sim E_c$, manifesting as a Gregory-Laflamme-like instability towards the minus branch of the black pole.  
This instability changes the topology of the horizon, resulting in a smaller horizon that covers only a part of the S$^3$: a black pole, as illustrated in Fig.\ref{fig:BHSol}. 

At the transition point, the black pole has a S$^3$ split into two regions with a horizon region and a region where the $y$ circle degenerates. The critical angle $\theta_c$ that delimits both regions on the S$^3$, \eqref{eq:AngleRanges} (and see Fig.\ref{fig:BHSol}),\footnote{The smooth region covers the S$^3$ from $\theta=0$ to $\theta_c$, while the black region ranges from $\theta=\theta_c$ to $\pi/2$.} satisfies $\cos^2 \theta_c= \varphi^{-1}$, where $\varphi=\frac{\sqrt{5}+1}{2}$ is the golden ratio. As we go to lower energies, one might expect that the black region would continue its fragmentation into a black belt, then a black bi-pole, etc. However, those saddles remain less dominant. Despite that, the black region over the S$^3$ still contracts as the energy decreases, so that the critical angle between the black region and the smooth region,
\begin{equation}
    \cos^2 \theta_c \= \frac{3-\sqrt{1-\frac{E}{E_\text{max}}}}{4}\,,
\end{equation}
reaches a minimum at $E=0$, where $\theta_c=\pi/4$ and the black region of the black pole covers exactly half of the S$^3$.

\subsection{Canonical ensemble}

In the canonical ensemble, the temperature, $T$, is fixed, while the energy can vary. The appropriate thermodynamic quantity for comparing different phases is the free energy $F(T)$. The free energy of the BTZ black hole as a function of temperature is given in \eqref{eq:FreeEnergyBTZ}, while the free energies of the black pole, black belt, and black bi-pole are given in \eqref{eq:LocBHEnEntFreeEn}, \eqref{eq:BlackBeltEnEntFreeEn} and \eqref{eq:DoubBHEnEntFreeEn} respectively. 

\begin{figure}[t]
    \centering
    \includegraphics[scale=1]{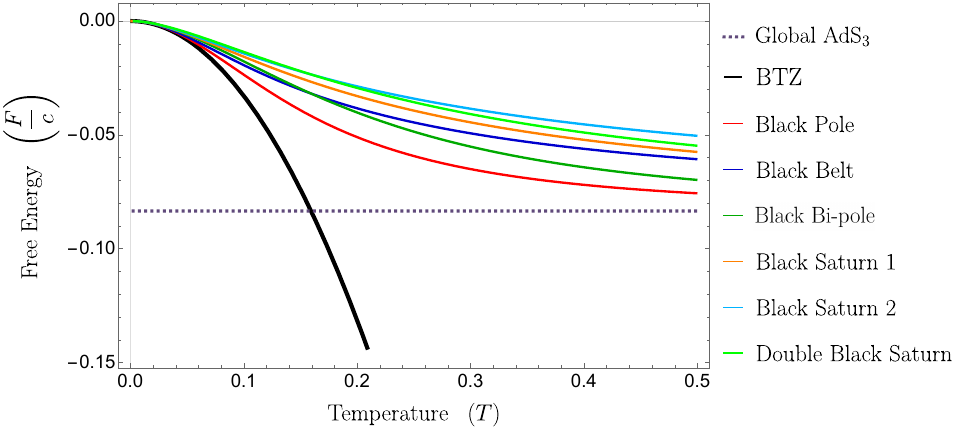}
    \caption{Free energy as a function of temperature for the BTZ black hole, global AdS$_3$, and several localized black hole geometries in the canonical ensemble. The most dominant phases are those predicted by Einstein gravity in three dimensions (Fig.\ref{fig:CFTbootstrap_F}), and all our new string-theory solutions have larger free energies.}
    \label{fig:canonical ensemble}
\end{figure}

Figure \ref{fig:canonical ensemble} plots the free energies of the BTZ black hole, global AdS$_3$ (Thermal AdS), and various localized black hole geometries in the canonical ensemble.

The plot shows that the free energies of all localized black hole geometries converge toward that of the zero-energy singular BTZ black hole as $T \to 0$ and approach global AdS$_3$ at high temperatures, interpolating between the two regimes. Importantly, they have larger free energies than both the BTZ black hole and global AdS$_3$ across the whole temperature range.

Thus, the thermodynamics in the canonical ensemble remains unaffected by our new ten-dimensional solutions, aligning instead with the predictions from pure Einstein gravity discussed in Section \ref{sec:EinsteinGrav}: Thermal AdS$_3$ dominates the ensemble up to $T = (2\pi)^{-1}$, with a transition to the BTZ black hole phase at higher temperatures.

The fact that the localized black hole geometries are missed in the canonical ensemble is also a confirmation of the expectations of  \cite{Banks:1998dd}. The nonequivalence of the phase structures between the canonical ensemble and the microcanonical ensemble is a direct consequence of the non-concavity of the entropy function $S(E)$ \cite{gross2001microcanonical}.
As seen in Fig.\ref{fig:NegEn} and Fig.\ref{fig:microcanonical}, the highest values of entropy follow a convex curve up to a certain energy threshold. This explains why the black pole dominates the microcanonical ensemble at low and medium energies $(-c/12<E<E_c)$,  while these energy ranges are absent in the canonical ensemble. Consequently, the canonical ensemble does not capture the transition from the BTZ phase to the black pole at $E = E_c$.

\section{Entanglement entropy and the shadow}
\label{sec:EEGen}

In the previous sections, we have demonstrated that new asymptotically-AdS$_3$ static black-hole solutions exist in supergravity on S$^3\times$T$^4$. These solutions have a nontrivial dependence of the S$^3$ azimuthal angle (in Hopf coordinates), and are not visible in pure Einstein gravity in three dimensions.
Notably, we have shown that these phases are more dominant than the BTZ black hole and global AdS$_3$ in the microcanonical ensemble for the low-energy and a part of the medium-energy range.

In this section, we aim to derive the microscopic properties associated with the localized black hole geometries appearing in the new phases. Specifically, we focus on the entanglement entropy, which can be found on the gravity side by computing the area of a minimal Ryu-Takayanagi (RT) surface\cite{RT06,RT06b}. After briefly reviewing the entanglement entropy in general quantum systems and the RT formula in holography, we propose a method to compute it in intricate ten-dimensional geometries that do not preserve the isometries of the internal manifolds. We calculate the area of the RT surfaces for both the BTZ black hole and the black pole. 

A black hole geometry can be interpreted either as a mixed state, such as a canonical or microcanonical ensemble, as we analyzed in the previous sections, or as a typical microstate (i.e., a pure state). The rules for RT computations differ between these two interpretations, and the details are reviewed in Appendix \ref{app:pure_mixed}.

In this section, we consider the holographic entanglement entropy of microstates associated with the static BTZ geometry and the black pole geometry. As we will see below, the holographic entanglement entropy, given by the {\it minimal} RT surface, is insensitive to the details of our new geometries. To capture these one must use non-minimal {\it extremal} RT surfaces, which go much deeper in the bulk, and which reveal significant differences between BTZ black holes and our new geometries. 

Previous calculations of the holographic entanglement entropy in AdS$_3$($\times {\rm S}^3 \times {\rm T}^4$) have focused mostly on solutions with many isometries, such as global AdS$_3$ (dual to the NS-NS ground state) with potential conical defects or the BTZ black hole \cite{RT06,RT06b,RT17}. The holographic entanglement entropy has also been calculated in some smooth and horizonless geometries that do not have charges corresponding to black holes with a macroscopic event horizon \cite{Giusto:2014aba,Giusto:2015dfa,Jones:2016iwx,Michel:2018yta,Bombini:2019vuk}.  In all of these examples, there exists a nontrivial isometry along the compact directions, and most of the time the RT surface is computed only close to the AdS boundary. 

We give the first example of a holographic entanglement entropy computation when the compact dimensions are nontrivially fibered on and mixed up with the AdS directions and the RT surface depends nontrivially on the coordinates of the internal space. Furthermore, our procedure allows us to estimate the area of RT surfaces that go deep in the infrared, and to compare the holographic entanglement entropies of distinct geometric states with the same energy. 

\subsection{Entanglement entropy and the Ryu-Takayanagi formula}\label{sec:RTForma}

Entanglement is a key property that characterizes quantum systems and is quantified by the entanglement entropy. The entanglement entropy turns out to be useful for understanding quantum many-body systems \cite{Bombelli:1986rw,Srednicki:1993im,Vidal:2002rm,Latorre:2003kg,Kitaev:2005dm,Calabrese:2004eu,Casini:2009sr} and spacetime emergence in the context of holography \cite{RT06,RT06b}.

Consider a quantum system that consists of two subsystems $A$ and $B$. The Hilbert space is given by a tensor product, $\mathcal{H}_{tot} = \mathcal{H}_A \otimes \mathcal{H}_B$,
where $\mathcal{H}_{A}$ (resp. $\mathcal{H}_{A}$) is the Hilbert space associated with the subsystem $A$ (resp. $B$). A pure state $\ket{\psi} _{tot}\in \mathcal{H}_{tot}$ can be decomposed into the general form:
\begin{align}
    \ket{\psi}_{tot} = \sum_i \lambda_i \ket{\alpha_i}_A \otimes \ket{\beta_i}_B, \qquad \text{when }\ket{\alpha_i}_A \in \mathcal{H}_A, ~\ket{\beta_i}_B \in \mathcal{H}_B. 
\end{align}
The state is said separable if there exist $\ket{\alpha}_A \in \mathcal{H}_A$ and $\ket{\beta}_B \in \mathcal{H}_B$ such that $\ket{\psi}_{tot} = \ket{\alpha}_A \otimes \ket{\beta}_B$. If no such decomposition exists, then $\ket{\psi}_{tot}$ is entangled. The amount of entanglement can be quantified by the entanglement entropy defined as the von Neumann entropy of the reduced density matrix $\rho_A$ of the subsystem $A$
\begin{align}
    S_{EE} = -{\rm Tr}\left(\rho_A \log \rho_A\right)\,,\qquad \rho_A = {\rm Tr}_B \left(|\psi\rangle_{tot}\langle \psi|_{tot}\right)\,.
\end{align}
Note that, for pure states,  we can alternatively use $\rho_B$ to get the same answer
\begin{align}\label{eq:symmEnt}
    -{\rm Tr}\left(\rho_A \log \rho_A\right) \equi -{\rm Tr}\left(\rho_B \log \rho_B\right).
\end{align}

The entanglement entropy reflects the fine-grained nature of quantum systems, but it is in general hard to compute in many-body systems and QFTs \cite{Calabrese:2004eu,Casini:2009sr,Calabrese:2009qy}. However, for those admitting semiclassical gravity duals, the entanglement entropy can be computed much more easily using the RT formula \cite{RT06,RT06b}. The RT formula states that the entanglement entropy of a spatial region $\cL$ in the field theory is given by the minimal codimension-2 spatial surface in the gravity side which is homologous to $\cL$ at the boundary. In its simplest form, the formula applies to theories whose gravity dual is classical Einstein gravity (eventually plus matter) and to states dual to static classical spacetimes that tend asymptotically to AdS$_d$. \\ 

\begin{figure}[t]
    \centering
    \includegraphics[width=\textwidth]{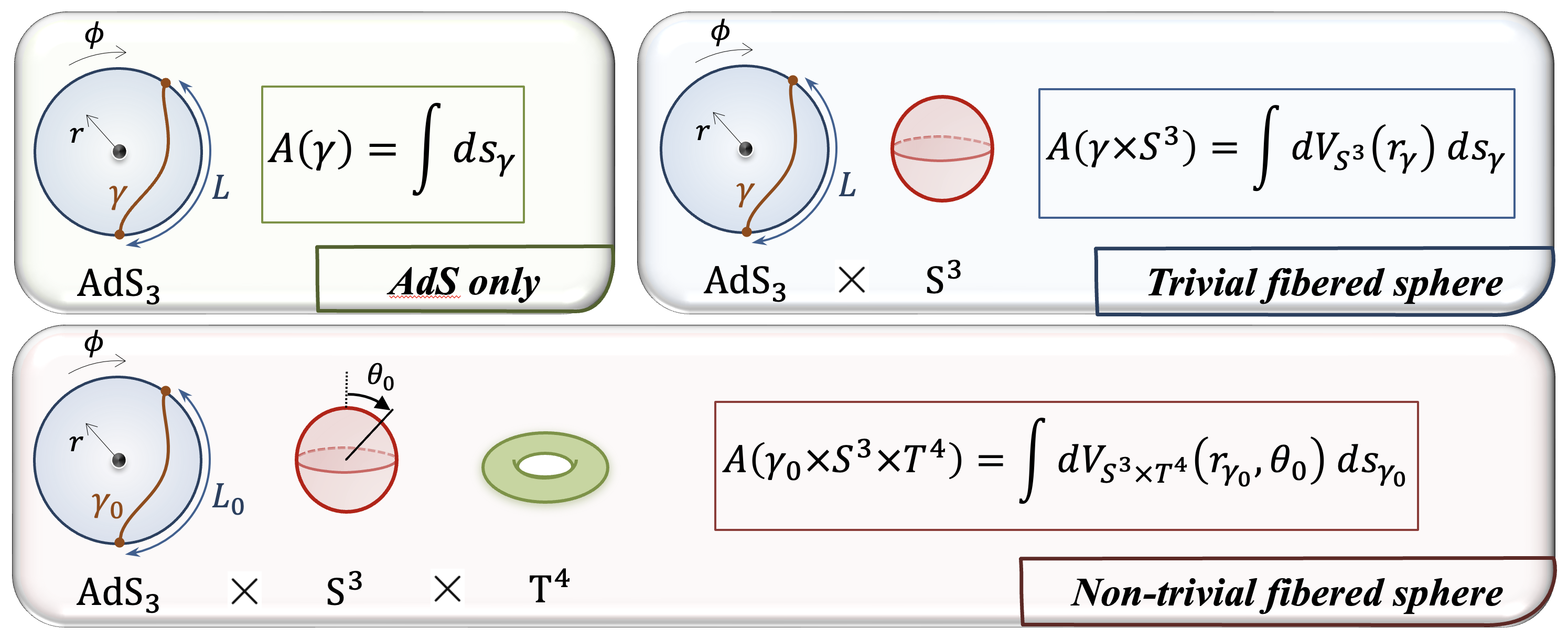}
    \caption{Schematic description of the RT formula and the RT surfaces for different geometries with AdS$_3$ asymptotics. Top left: original formulation \cite{RT06,RT06b} in AdS$_3$. Top right: extension to asymptotically-AdS$_3\times$S$^3$ backgrounds with an almost-product structure\cite{Giusto:2014aba,Bombini:2019vuk}. Bottom: extension to our ten-dimensional geometries that correspond to nontrivial S$^3\times$T$^4$ fibrations over AdS$_3$ by finding $\gamma_0$ with a fixed position on the S$^3$, $\theta=\theta_0$. }
    \label{fig:EEplots}
\end{figure}

In AdS$_3$, the boundary CFT describes a critical quantum system defined on an S$^1$, parametrized by $\phi \equiv y/R_y$ with a total circumference of $2\pi$ in our convention. For $y$-independent geometries, a connected one-dimensional spatial region $\cL$ at the boundary is uniquely specified by its length $L$. The holographic entanglement entropy for such $\cL$ is given by the RT formula
\begin{equation}
    S_{EE}(L) \= \frac{1}{4G_3}\,\underset{\partial \gamma = \partial \cL}{\min} A(\gamma)\,, \qquad A(\gamma) \equi \int ds_\gamma\,, 
\end{equation}
where $\gamma$ is a curve which ends at the boundary on $\partial \cL$, $G_3$ is the three-dimensional Newton constant, $A(\gamma)$ is the length of $\gamma$, and $ ds_\gamma$ is the line element along $\gamma$ (see the top left of Fig.\ref{fig:EEplots}). 

To apply the RT formula to the D1-D5 CFT, it must be adapted to static geometries that asymptote to AdS$_3\times\cM$ where $\cM$ is a compact space. The original RT paper \cite{RT06b} only considered the situation where the compact space forms a trivial direct product with the AdS directions. In \cite{Giusto:2014aba,Bombini:2019vuk}, the authors performed a computation in warped AdS$_3\times$S$^3$  geometries that have an almost-product structure, and where the RT surface wraps the internal ($\rm S^3 \times T^4$) space. The RT formula is:
\begin{align}
    S_{EE}(L) \= \frac{1}{4G_6}\,\underset{\partial \Gamma_4 = \partial \cL \times S^3}{\min} A(\Gamma_4)\,, \qquad A(\Gamma_4)\equi \int dV_{S^3}(r_\gamma) \, ds_\gamma\,,
\end{align}
where $\Gamma_4=\gamma\times S^3$ is the warped product between a curve $\gamma$ in AdS$_3$ and the S$^3$, and $dV_{S^3}(r_\gamma)=\sqrt{\det g|_{S^3}(r_\gamma)} \,\,d\theta d\varphi_1 d\varphi_2$ is the volume form of the S$^3$ along the curve $\gamma$ in AdS$_3$ (see the top right of Fig.\ref{fig:EEplots}).

This allows the RT formula to be reduced to an effective three-dimensional problem:
\begin{equation}
    A(\gamma \times S^3) \= 2\pi^2 \, \int R_{S^3}(r_\gamma)^3  ds_\gamma\,,
    \label{eq:AreaGeoInS3}
\end{equation}
where $R_{S^3}(r_\gamma)$ is the S$^3$ radius along the curve, with $\sqrt{\det g|_{S^3}(r_\gamma)}=R_{S^3}(r_\gamma)^3 \cos\theta \sin \theta$. Hence, for computing the area of the RT surface one can consider an effective three-dimensional metric $d\widetilde{s}_3^2 \= R_{S^3}(r_\gamma)^6\, ds_3^2$, accounting for the S$^3$ deformation and reducing the problem to the original RT formula. \\

The localized black hole geometries we construct in this paper depend non-trivially on  $\theta$, the Hopf polar angle of the S$^3$, and hence do not fit in the class of solutions described above. 
This complicates the application of the RT formula, as finding a minimal codimension-2 surface ending at the boundary on $\partial \cL \times$S$^3\times$T$^4$ no longer corresponds to finding a geodesic curve in the AdS$_3$ part. Instead, the minimal surface depends on two parameters: an affine parameter $\lambda \in [0,1]$, and the angle $\theta$; its embedding is given by a curve $r(\lambda,\theta)$. Consequently, at a given $\lambda_0\neq 0,1$, the minimal surface spans a range of radii in AdS$_3$, $r(\lambda_0,\theta)$, for $0\leq \theta\leq \pi/2$. 

Our method restricts consideration to a subset of codimension-2 surfaces corresponding to curves within the AdS$_3$ region. We then perform a minimization computation for such a subset of codimension-2 surfaces and estimate the entanglement entropy. The recipe decomposes into the following steps:

\begin{itemize}
    \item[1-] \underline{\textbf{Finding curves in AdS$_3$:}}
\end{itemize}
First, we aim to find geodesic curves in the AdS$_3$ part, homologous to $\cL$ at the boundary, that also take into account the deformation of the S$^3\times$T$^4$. 
Specifically, we fix first the position on the S$^3$, $\theta=\theta_0$, where $\theta_0$ is a free parameter that we will vary, and identify the curve $\gamma_0$ ending on $\partial \cL$ at the boundary, which minimizes a quantity incorporating the S$^3\times$T$^4$ deformation akin to \eqref{eq:AreaGeoInS3}. More precisely, we derive the curve $\gamma_0:\lambda \in [0,1] \to (r(\lambda),\phi(\lambda))$ that minimizes
\begin{equation}
    \cA \= \int  \frac{\sqrt{\det g|_{S^3\times T^4,\,\theta=\theta_0}}}{\sin \theta_0 \,\cos \theta_0}\, ds_{\gamma_0, \,\theta=\theta_0},\quad  ds_{\gamma_0, \,\theta=\theta_0} \= \sqrt{g_{\mu \nu}(\theta=\theta_0) \frac{dx^\mu}{ d\lambda}\frac{dx^\nu}{ d\lambda}}\, d\lambda,\quad x^\mu=(r,y).
    \label{eq:GeodesDef}
\end{equation}
This provides a set of curves $\gamma_0$ labelled by $0\leq \theta_0 \leq \pi/2$.
\begin{itemize}
    \item[2-] \underline{\textbf{Derivation of the area of the codimension-2 surface along $\gamma_0$:}}
\end{itemize}
Subsequently, we compute the area of the codimension-2 surface along this curve in the whole ten-dimensional geometry (see bottom of Fig.\ref{fig:EEplots})
\begin{equation}
   A(\gamma_0 \times S^3 \times T^4) \equi (2\pi)^6 \, V_4 \, \int \,\sqrt{\det g|_{S^3\times T^4}} \,\, ds_{\gamma_0,\theta}\, d\theta \,,
   \label{eq:MinSurfTheta0}
\end{equation}

\begin{itemize}
    \item[3-] \underline{\textbf{Estimation of the entanglement entropy:}}
\end{itemize}
From the areas \eqref{eq:MinSurfTheta0}, we can determine the minimal value by varying $\theta_0$: $\underset{\theta_0}{\min} \,A(\gamma_0 \times S^3 \times T^4)$.
Since we have changed the order of minimization, it is not clear if this represents the RT surface for AdS$_3\times$S$^3\times$T$^4$ spacetimes with a nontrivial product structure. However, we necessarily have an upper bound $S_{EE}(L)\leq\frac{1}{4G_{10}}\underset{\theta_0}{\min} \,A(\gamma_0 \times S^3 \times T^4)$. Through detailed derivation, we cautiously propose 
\begin{equation}
S_{EE}(L) \,\approx\, \frac{1}{4G_{10}} \underset{\theta_0}{\min} \,A(\gamma_0 \times S^3 \times T^4).
    \label{eq:EntanglementEntropLocBH}
\end{equation} \\

To derive the quantities \eqref{eq:GeodesDef} and \eqref{eq:MinSurfTheta0}, it is also relevant to define an effective three-dimensional metric incorporating the deformation factor $\sqrt{\det g|_{S^3\times T^4}}$:
\begin{equation}
    d\widetilde{s}_3^2 \equi R_{S^3\times T^4}(r,\theta)^6 \,  ds_3^2\,,\qquad \text{where  }\det g|_{S^3\times T^4}\equi R_{S^3\times T^4}(r,\theta)^6\,\cos^2\theta\sin^2\theta\,.
\end{equation}
For our geometries \eqref{eq:metAdS3+BTZs},  this metric takes the form:\footnote{Note that we have to work with the ten-dimensional metric in the Einstein frame, $g^E_{\mu\nu}= e^{-\Phi/2} g_{\mu \nu}$.}
\begin{equation}
   d\widetilde{s}_3^2 \= Q_1 Q_5 \,G(r,\theta)  \left[-r^2 \,\cK_y(r,\theta) \,dt^2 +R_y^2\,\frac{r^2+\ell^2}{\cK_y(r,\theta)} \, d\phi^2 +Q_1 Q_5 \,G(r,\theta) \,\frac{ dr^2}{r^2+\ell^2} \right]\,.
   \label{eq:Eff3dMet}
\end{equation}
As such, finding $\gamma_0$ involves solving a geodesic problem within $d\widetilde{s}_3^2$, restricted to $\theta=\theta_0$, and minimizing the length
\begin{equation}
    \cA \= \int \sqrt{\widetilde{g}_{r r}(r,\theta_0) \left( \frac{ dr}{ d\lambda}\right)^2 +\widetilde{g}_{\phi\phi}(r,\theta_0) \left( \frac{ d\phi}{ d\lambda}\right)^2 }\, d\lambda\,,
\end{equation}
where $\gamma_0$ ends at the boundary on $\partial \cL$. This geodesic problem is governed by the equations:
\begin{equation}
    \frac{d}{ d\lambda}\left(\widetilde{g}_{\phi\phi}(r,\theta_0) \frac{ d\phi}{ d\lambda} \right) \= 0 \,, \qquad \widetilde{g}_{r r}(r,\theta_0) \left( \frac{ dr}{ d\lambda}\right)^2 +\widetilde{g}_{\phi\phi}(r,\theta_0) \left( \frac{ d\phi}{ d\lambda}\right)^2 \=1\,.
    \label{eq:GeodesEq}
\end{equation}
Once the curve $\gamma_0$ is determined, we can derive $A(\gamma_0 \times S^3 \times T^4) $ \eqref{eq:MinSurfTheta0}, by integrating:
\begin{equation}
    A(\gamma_0 \times S^3 \times T^4) \equi (2\pi)^6 \, V_4 \, \int \,\sqrt{\widetilde{g}_{r r}(r,\theta) \left( \frac{ dr}{ d\lambda}\right)^2 +\widetilde{g}_{\phi\phi}(r,\theta) \left( \frac{ d\phi}{ d\lambda}\right)^2 }\,\cos \theta \sin \theta \,  d\lambda d\theta.
    \label{eq:RTsurfaceGen}
\end{equation}
For a detailed resolution of this geodesic problem and an expression of $A(\gamma_0 \times S^3 \times T^4)$ as a function of $L$, the length of the boundary interval $\cL$ for which the entanglement entropy is derived, we refer interested readers to Appendix \ref{app:EELocBH}. \\

In the next subsections, we present an analytic computation of holographic entanglement entropy at the small-subsystem limit for the ten-dimensional geometries constructed in this paper. Then, we provide numerical results (and analytic results if available) on minimal RT surfaces and non-minimal RT surfaces for the black pole and the BTZ black hole, and discuss their differences.

\subsection{Small subsystems and the entanglement first law}

Deriving the RT surfaces of the intricate localized black-hole geometries generically requires numerical analysis. However, in the short-interval limit, $L\ll 1$, one can compute the area of these surfaces analytically (we remind the reader that $L$ is the length of the one-dimensional spatial region of the subsystem $\cL$).

We can do this for a generic solution from the class introduced in Section \ref{eq:LocBHSol}. By expanding the effective metric \eqref{eq:Eff3dMet} at large $r$, and solving the geodesic equations \eqref{eq:GeodesEq} near the boundary for small $L$, we derive the RT surfaces \eqref{eq:RTsurfaceGen}. Details of this derivation are provided in Appendix \ref{app:SmallGeo}. The resulting expression is:
\begin{equation}
     A(\gamma_0 \times S^3 \times T^4) \= (2\pi)^6 V_4 Q_1 Q_5 \left[\log\left(\frac{r_\infty\, L R_y}{\sqrt{Q_1Q_5}}\right)+\frac{L^2E}{2c} \right]
\end{equation}
where $r_\infty$ is the radial cutoff to regularize the divergence of geodesics reaching the boundary ($r_\infty\to \infty$), and $E$ is the energy of the geometry as derived in the bulk in \eqref{eq:Energyboundstate}. The bulk radial cutoff, $r_\infty$, corresponds to a lattice cutoff, $\epsilon$, on the boundary CFT side via the relation 
\begin{align}
    \epsilon \equiv \frac{\sqrt{Q_1 Q_5}}{r_{\infty} R_y} = \frac{R_{AdS}^2}{r_{\infty} R_y}.
\end{align}
Thus, the surfaces are independent of $\theta_0$, so that the minimization over the angular position on the S$^3$ is trivial and the entanglement entropy formula \eqref{eq:EntanglementEntropLocBH} trivially gives:
\begin{equation}
    S_{EE}(L)\,\sim\, \frac{c}{3}\log\left(\frac{L}{\epsilon}\right)+\frac{L^2E}{6} , \qquad L\ll 1\,.
    \label{eq:ShortIntervalEE}
\end{equation}
Note that the first logarithmic term is nothing but the entanglement entropy at $E=0$, and the second term is proportional to the energy density $E/2\pi$ and $L^2$. This behavior is known as the entanglement first law \cite{BNTU12,WKZV13,BCHM13,FGHMvR13,RT17,TT21}, which holds for any two-dimensional QFT in the short-interval limit.

Thus, for small $L$, there is no distinction between the entanglement entropy of a localized black hole geometry and that of a BTZ black hole with the same energy, as anticipated from the entanglement first law. A natural question then arises: does this equivalence hold across the entire range $L \leq \pi$, where the RT surface captures the entanglement entropy of finite subsystems, or do differences emerge?

\subsection{Finite subsystems and non-minimal RT surfaces} 

 In this section, we evaluate the entanglement entropy for subsystems with finite length, $L=\cO(1)$. 

\subsubsection{Entanglement entropy for the BTZ black hole}

The RT surfaces of the BTZ black hole, described in Type IIB supergravity in Section \ref{sec:BTZ}, can be derived analytically using the RT formula outlined in Section \ref{sec:RTForma}. The resulting expression for the entanglement entropy is:
\begin{equation}
 S_{EE\,BTZ}(L) \= \frac{c}{3} \log\left[\sqrt{\frac{c}{3E}}\,\frac{1}{\epsilon} \,\sinh \left(\sqrt{\frac{3E}{c}}\,L \right) \right],~~~~ 0 < L \leq \pi
    \label{eq:BTZEE}
\end{equation}
In the small-interval limit $L\ll 1$, this expression reduces to \eqref{eq:ShortIntervalEE}. However, we can get a similar expression from a ``small $E/c$'' expansion: 
\begin{align}
    S_{EE~BTZ}\,\sim\, \frac{c}{3}\log\left(\frac{L}{\epsilon}\right)+\frac{L^2E}{6} , \qquad E/c \ll 1\,.
    \label{eq:EEsmallE}
\end{align}
Thus, at low energies, the entanglement entropy reduces to a universal value only determined by the energy, even at finite subsystem length $L$.

Since we aim to compare the BTZ black hole and localized black holes, especially the black pole, we consider the range $0<E \leq c/96$ where both of them coexist. In this energy range, the entanglement entropy is remarkably approximated by \eqref{eq:EEsmallE}, even for the largest $L\sim \pi$.\footnote{In the energy range considered, $\sqrt{\frac{E}{c}}\lesssim \frac{1}{10}$. Therefore, the deviation from \eqref{eq:EEsmallE}, of order $L^4(E/c)^2$, is roughly $ 10^{-2} $ even for $L\sim \pi$.} 

Note that if we had instead focused on a range closer to the Hawking-Page transition energy, $E\sim c/12$, the entanglement entropy would have captured details beyond the universal ``small $E/c$'' limit. Indeed, in that scenario, $L\sqrt{\frac{3E}{c}}$ is of order 1 for $L\sim \pi$ so that next-order terms in \eqref{eq:EEsmallE} contribute significantly. The black pole and other localized black hole geometries start appearing at $E\sim c/100$, and hence they are in the ideal energy regime to ``hide'' their contribution to the entanglement entropy behind the BTZ universal ``small $E/c$'' limit  \eqref{eq:ShortIntervalEE}.

\subsubsection{Non-minimal RT surfaces in the BTZ black hole}
\begin{figure}[t]
    \centering
    \includegraphics[width=0.82 \textwidth]{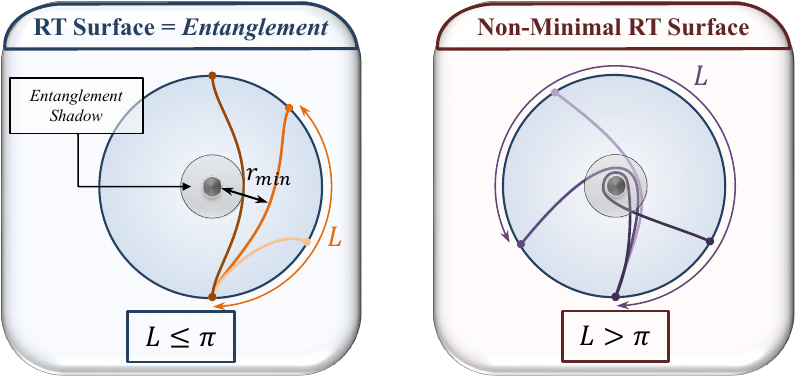}
    \caption{The figure illustrates extremal RT surfaces as a function of $L$.  When $L\leq \pi$ (left-hand side), the RT surface is minimal, and gives the entanglement entropy of the boundary region defined of length $L$. For $L> \pi$ (right-hand side), the RT surfaces represent extremal surfaces that are not minimal, and are believed to probe the internal degrees of freedom on the CFT side.}
    \label{fig:EntangEntwin}
\end{figure}

 The expression \eqref{eq:BTZEE} corresponds to the entanglement entropy only for $0<L\leq \pi$. On the other hand, if one extends this expression to  $L>\pi$, it represents the area of non-minimal extremal surfaces, as shown in Fig.\ref{fig:EntangEntwin}. We will call these surfaces non-minimal RT surfaces. As sketched in Fig.\ref{fig:EntangEntwin}, the union of all minimal RT surfaces does not cover the entire spacetime, leading to an \emph{entanglement shadow} in the infrared region. The BTZ black holes exhibit an entanglement shadow approximately the size of the AdS scale around the horizon \cite{Hubeny:2013gta,Balasubramanian:2014sra}.
As we increase $L>\pi$, the non-minimal RT surface wraps closer and closer to the black-hole horizon, and protrudes insight into the entanglement shadow.

The CFT interpretation of non-minimal RT surfaces is not precisely known. Various arguments have suggested that the non-minimal RT surfaces may probe the internal degrees of freedom on the CFT side \cite{Balasubramanian:2014sra,Caminiti:2024ctd,Arora:2024edk}.

For the BTZ black hole, the non-minimal RT surface starts probing internal degrees of freedom captured by the Bekenstein-Hawking entropy of the geometry. Indeed, at large $L$, it exhibits linear growth in $L$: 
\begin{equation}
    \frac{c}{3} \log\left[\sqrt{\frac{c}{3E}}\,\frac{1}{\epsilon} \,\sinh \left(\sqrt{\frac{3E}{c}}\,L \right) \right] \,\sim\, S_\text{BTZ}\,\,\frac{L}{2\pi}\,,\qquad \text{when }L \gg \sqrt{\frac{c}{E}}\,,
    \label{eq:EntropyGrowthBTZ}
\end{equation}
where $S_\text{BTZ}$ is the BTZ entropy \eqref{eq:BTZEntropvsEnergy}. This linear increase in terms of the black hole entropy occurs because large-$L$ RT surfaces protrude inside the BTZ entanglement shadow and encircle the black hole horizon. Hence, their area grows in length proportional to how many times they cover the area of the horizon.

 This BTZ derivation suggests that for any geometries in the medium energy range with $0 \leq E\lesssim c/100$, the entanglement entropy will follow the same universal expression as for the short-interval limit \eqref{eq:ShortIntervalEE} since we have $L \ll \sqrt{\frac{c}{E}}$ for $L\leq \pi$. For larger $L$, the non-minimal RT surfaces begin to capture the inner details of the geometry, which can vary significantly depending on the specific background.

\subsubsection{The entanglement entropy in the black-pole geometry}

We derive the RT surfaces in the black-pole geometry as a function of $L$ using the formalism detailed in Section \ref{sec:RTForma} and Appendix \ref{app:EELocBH}. Our aim is to compare the area of these surfaces with those of a BTZ black hole with the same energy. However, unlike in the BTZ geometry, the minimal surface problem for black poles can only be solved numerically. 

As discussed in Section \ref{sec:MicroCano}, we recall that at fixed energy between $0$ and $E_\text{max}=c/96$, there are two black-pole solutions: a plus branch, which is less entropic than the BTZ black hole and a minus branch which is more entropic (see Fig.\ref{fig:microcanonical}).

We focus on the minus branch, which dominates the microcanonical ensemble at $-c/12 < E\leq c\left(5\sqrt{5}-11\right)/24 \sim 0.0075 c$, and later briefly discuss other localized black hole geometries.

\begin{figure}[t!]
    \centering
    \includegraphics[width=\textwidth]{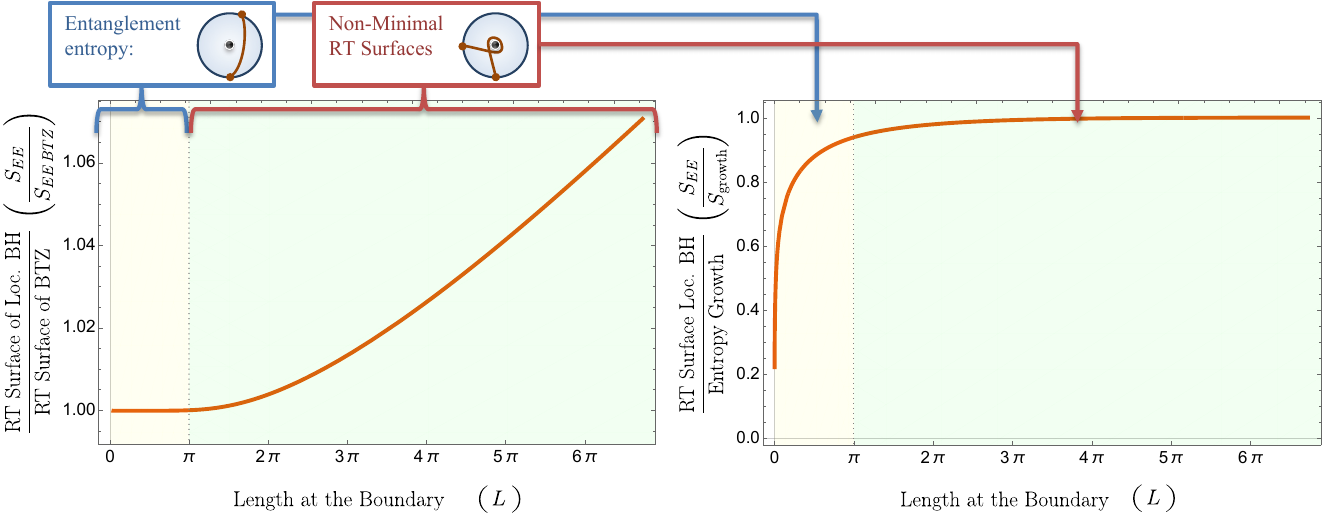}
    \caption{RT surfaces of a black pole with $E=10^{-3} c$ as a function of the length $L$ of the subsystem at the AdS$_3$ boundary. Left plot: ratio of the black-pole RT surface area to that of a BTZ black hole with the same energy (and less entropy). Right plot: ratio of the surface area to the entropy growth function \eqref{eq:EntropyGrowthBP}. Both plots are divided into two regions: for $0\leq L\leq \pi$, the RT surface is minimal and captures the entanglement entropy;  for $L>\pi$, the RT surface is non-minimal and probes the internal degrees of freedom in the corresponding CFT.}
    \label{fig:EEBlackPole}
\end{figure}

By deriving RT surfaces for various black poles with different energies, we observed that the energy does not significantly influence the main features of their RT surfaces. In Fig.\ref{fig:EEBlackPole}, we have plotted the RT surfaces as a function of $L$ for representative black pole with energy $E=10^{-3}c$ (which has more entropy than the BTZ black hole with the same energy; see Fig.\ref{fig:microcanonical}).

The plot on the left shows the ratio of the RT surface area of the black pole to that of the BTZ black hole with the same energy \eqref{eq:BTZEE}. It is evident that both functions match up to $L \sim \pi$, the regime where the areas capture the entanglement entropy. This confirms our expectation that any geometry at sufficiently low energies will exhibit the same universal entanglement entropy, which depends only on the central charge, $c$, the boundary size, $R_y$, and the energy, $E$. 

This result shows that the entanglement entropy of any interval at the boundary is identical whether we consider a pure state described by the BTZ black hole, or a pure state described by the black pole.

Our construction does not explain why localized black holes exist only below $c/96$, but this range is ideal for matching the entanglement entropy of the BTZ black hole, the dominant saddle in the canonical ensemble. If we could find other localized black holes (using for example the technology of \cite{Dias-Santos-toappear}) at relatively higher energies, of order $E\sim c/12$, it would undermine the universality of the entanglement entropy and the shadow conjecture we put forth in the Introduction.

\subsubsection{Non-minimal RT surfaces in the black pole}

Beyond $L>\pi$, the non-minimal RT surface areas start to deviate from those of the BTZ black hole as they begin to enter the entanglement shadow and probe the inner structure of the geometry, which significantly differs from the BTZ black hole. The plot on the right-hand side of Fig.\ref{fig:EEBlackPole} shows the ratio of RT surfaces in the black pole geometry to the entropy growth expected for any geodesic encircling the horizon of a black object. This entropy growth is similar to that of the BTZ black hole \eqref{eq:EntropyGrowthBTZ}, with the black pole entropy:\footnote{For the plot in Fig.\ref{fig:EEBlackPole}, we have also added a constant term in $S_\text{growth}$, which depends on $\log 1/\epsilon$, a large value: $S_\text{growth}+\frac{c}{3} \log \left( \sqrt{\frac{c}{3E}}\,\frac{1}{\epsilon}\right) .$}
\begin{equation}
    S_\text{growth} \equi S^-_\text{BP}\,\,\frac{L}{2\pi}\,,
    \label{eq:EntropyGrowthBP}
\end{equation}
where $S^-_\text{BP}$ is the entropy of the minus branch given in \eqref{eq:BlackPoleEntropIn}. The plot clearly shows that soon after the RT surface enters the entanglement shadow it starts longing the event horizon, and its area strongly grows like the entropy multiplied by the number of times the RT surface wraps around the horizon. 

Thus, our analysis of the entanglement entropy and the entanglement shadow has illustrated what appears to be a general characteristic of regular geometries in AdS$_3$ at low energies. The entanglement entropy is insensitive to the different geometries and is governed by the universal behavior obtained from the small-energy limit \eqref{eq:EEsmallE}, or equivalently the short-interval limit \eqref{eq:ShortIntervalEE}. We have verified this statement by deriving the RT surfaces of other localized black hole geometries: the black pole of the plus branch, the black belt, and the black bi-poles. In all these solutions, the RT surfaces exhibit the same features as the RT surface of the black pole, shown in Fig.\ref{fig:EEBlackPole}. The areas match those of the BTZ black hole with the same energy up to a certain boundary length, and they start matching the entropy growth function for larger $L$. 

The differences between these geometries in terms of RT surfaces are visible when $L>\pi$. The first difference is in the boundary length at which the non-minimal RT surfaces start to deviate from those of the BTZ black hole. This length can be much larger than $L\sim \pi $, especially for the plus branch of the localized black hole geometries, which have more BTZ-like characteristics than the minus branch. The second difference is whether the non-minimal RT surface areas become larger (as in Fig.\ref{fig:EEBlackPole}) or smaller than those of the BTZ black hole. This depends on the entropy of the geometry compared to that of the BTZ black hole. \\

Non-minimal RT surfaces probe the region of the entanglement shadow, where the mixing between the compact directions and the AdS directions becomes significant. Since the compact directions correspond to the internal degrees of freedom of the D1-D5 CFT, our results suggest that microstates described by different black geometries with the same energy have a very different organization of the internal degrees of freedom, and non-minimal RT surfaces are sensible to them. 

This is consistent with previous studies on non-minimal RT surfaces. For example, it is found in \cite{Balasubramanian:2014sra} that, in some special examples, the area of a non-minimal RT surface coincides with the entanglement entropy between two parties of internal degrees of freedom (but not between two spatial regions) on the CFT side. Besides, the authors of \cite{Caminiti:2024ctd} have recently found that the bulk scattering process, which heavily relies on the sub-AdS locality and is hence related to the internal degrees of freedom \cite{Heemskerk:2009pn}, is strongly constrained by the area of certain non-minimal RT surfaces.

\section{Conclusion and discussion}
\label{sec:Conclusion}

In this section, we present a concise summary of the results and outline a non-exhaustive list of potential future directions.

\subsection{Summary of results}

We have constructed a class of asymptotically ${\rm AdS}_3\times {\rm S}^3\times {\rm T}^4$ localized black holes, including configurations such as the black pole, the black belt, black bi-pole, and more fragmented geometries. These non-supersymmetric black holes exist in the energy range $-\frac{c}{12}< E < \frac{c}{96}$. This range lies within what we defined as the canonical shadow, the energy range where universal constraints of large-$c$ CFT fail to capture the microcanonical ensemble.

We have analyzed the thermodynamic properties of these geometries in detail. Specifically, in the range $-\frac{c}{12}<E< \frac{c}{24}\left(5\sqrt{5}-11\right)$, the black pole exhibits the highest entropy among the localized black holes, making it the dominant solution in the microcanonical ensemble. For higher energies, the BTZ geometry dominates over all the localized black holes. In the canonical ensemble, however, the behavior is different: the dominant configurations are either the thermal AdS geometry or the BTZ black hole, with a Hawking-Page transition occurring at $T=1/2\pi$, similar to what happens in pure AdS$_3$ Einstein gravity. The localized black holes remain subdominant in this ensemble. These features of the thermodynamics are consistent with the properties of the dual CFT$_2$ inferred from the large-$c$ modular bootstrap. Moreover, our analysis provides insights beyond the modular bootstrap, which becomes less predictive below $E=c/12$. 

We have computed the holographic entanglement entropy for the localized black hole geometries, focusing on the black pole. The RT surface is an eight-dimensional extremal surface in the bulk, which depends nontrivially both on the AdS coordinates and on the azimuthal angle on the $\rm S^3$. We proposed a recipe to estimate for the first time the area of such surfaces, which can be used in other solutions where the AdS metric depends non-trivially on the compact internal directions. The analytic results for the small subsystems match, as expected the entanglement first law. 

The surprise occurs for larger subsystems, where the area of the RT surface was computed numerically: for geometries that differ drastically but have the same energy (such as a black pole and a BTZ black hole) the holographic entanglement entropies are approximately identical. The root of this surprising agreement is the fact that the localized black holes only exist below $E = c/96$, a regime where the universal entanglement first law holds for boundary intervals of non-negligible length. From a geometric perspective, this implies that the differences between our new solutions and the BTZ black hole are restricted to its entanglement shadow, and hence are not captured by minimal RT surfaces. 

However, these explanations introduce a more profound question: what fundamental mechanism restricts the existence of localized black holes to this energy range or, equivalently, restricts the differences between the localized black holes and the large black hole to lie inside the latter's entanglement shadow. Finding this fundamental mechanism would allow one to prove the {\em Shadow Conjecture} we have formulated in the Introduction. 

On the other hand, we found that non-minimal RT surfaces, which probe the entanglement shadow and capture internal degrees of freedom, exhibit clear distinctions across different geometries. This aligns with the expectation that non-minimal RT surfaces are sensitive to the fine-grained structure of the spacetime at sub-AdS scales.

Despite uncovering many intriguing features of localized black hole geometries, their connection to the physics of the dual CFT remains incomplete. Below, we outline several potential connections and directions for future investigation.

\subsection{Some future directions}

\subsubsection{Estimation of the HEE for non-static geometries in AdS$_3\times$S$^3\times$T$^4$}

In Section \ref{sec:RTForma}, we introduced a procedure to estimate the holographic entanglement entropy for states dual to geometries with AdS$_3\times$S$^3\times$T$^4$ asymtpotics that do not have direct-product structure, by generalizing the standard RT computation to static spacetimes with nontrivial dependence on the S$^3$ coordinates.

Extending this procedure to non-static geometries would be a significant advancement. For non-static AdS spacetimes without internal directions, the entanglement entropy can be computed with the Hubeny-Rangamani-Takayanagi (HRT) formula \cite{Hubeny:2007xt}. Similarly, this extension would involve adapting the HRT formalism to AdS$_3\times\cM$ solutions with nontrivial dependence along the $\cM$ directions.

Such a generalization is crucial for testing the Shadow Conjecture, which requires calculating the entanglement entropy for various black geometries distinct from BTZ. These geometries often involve momenta in AdS$_3$ and angular momenta along the S$^3$, providing a rich landscape for investigation.

\subsubsection{New energy thresholds from the CFT}

In pure Einstein gravity in three dimensions, the key energy threshold occurs at $E = 0$, marking the emergence of the BTZ black hole. As reviewed in Section \ref{sec:2}, this threshold has no direct counterpart in large-$c$ CFTs. Our ten-dimensional solutions confirm that there is nothing inherently special about $E = 0$. Instead, this apparent threshold arises from the truncated nature of pure Einstein gravity, which excludes essential field content from string theory that is crucial for capturing small black hole geometries. Our constructions introduce two new energy thresholds, $E = c/96$ and $E = (5\sqrt{5} - 11)c/24$, raising the same question of whether these thresholds have CFT interpretations.

The energy $E = c/96$ corresponds to the threshold below which localized black holes begin to exist. Understanding how the factor $1/96$ emerges from the CFT perspective would be highly interesting. Notably, $96 = 12 \times 8$, and the factor $1/12$ appears prominently in CFT$_2$ calculations, from conformal anomalies to modular bootstrap. Similarly, $96 = 16 \times 6$, and the factor $1/16$ plays an important role in AdS$_3$/CFT$_2$. It appears in the large-spin bootstrap \cite{Kusuki:2018wpa, Collier:2018exn} and in the conformal bootstrap for boundary CFTs \cite{Kusuki:2022wns, Kusuki:2022ozk}. While it could be coincidental that both $12$ and $16$ are factors of $96$, this observation could suggest that $E=c/96$ should admit a CFT counterpart.

Another intriguing value is the threshold $E = (5\sqrt{5} - 11)c/24$, where a phase transition occurs between the BTZ geometry and the black pole geometry. As discussed in Section \ref{sec:MicroCano}, the golden ratio emerges as the angular length of the horizon on the $\rm S^3$ at this specific energy. While this is intriguing, we currently lack a deeper understanding of why this happens. Gaining insight into how this constant arises from the CFT perspective would be a significant step forward.

\subsubsection{Testing the shadow conjecture}

The localized black holes discussed in this paper all reside within the ``canonical shadow," defined as the energy range $-c/12 < E < c/12$, which cannot be accessed by the canonical ensemble. However, we have found that these localized black holes are further restricted to the narrower range $-c/12 < E < c/96$. At first glance, the exclusion of the $c/96 < E < c/12$ window might appear coincidental. However, this range ensures that the localized black holes and BTZ black holes differ only inside their entanglement shadows as shown in Section \ref{sec:EEGen}. Consequently, the entanglement entropies of the BTZ black hole and the localized black holes are approximately identical, governed by the universal entanglement first law. This overlap of the canonical and entanglement shadows motivated us to propose the {\em Shadow Conjecture} in the Introduction.

It would be valuable to further investigate this connection on the gravity side. This involves constructing additional small black hole solutions in AdS$_3$, focusing on the canonical shadow energy range $-c/12 < E < c/12$ and on solutions dominating the microcanonical ensemble. The small black holes presented here arise from a restrictive integrable ansatz in Type II supergravity, which might not capture the full spectrum of potential solutions. For example, our ansatz constrains all localized black holes to have an S$^3 \times$S$^1$ or S$^2 \times$T$^2$ horizon topology, excluding the possibility of small black holes with an S$^4$ horizon topology, which cannot be constructed within our framework but can be constructed numerically \cite{Dias-Santos-toappear}, following an approach similar to \cite{Dias:2016eto}. Additionally, the solutions presented in this paper, obtained from \cite{Bah:2022pdn}, rely on a restrictive linear branch of the equations. 

It would be interesting to exhaust the space of localized black-hole solutions with $\rm AdS_3$ asymptotics, and to verify whether all microcanonically dominant phases live inside the BTZ entanglement shadow. It would also be important to verify the Shadow Conjectures for the localized non-extremal black holes with $\rm AdS_5$ \cite{Dias:2015pda, Dias:2016eto}, or with ${\rm AdS}_{\text{large-}d}$ asymptotics \cite{Herzog:2017qwp}, and to construct other nonextremal solutions where they could be tested (using for example the blackfold approach \cite{Emparan:2009at}). It would also be interesting to test the Shadow Conjecture for supersymmetric black holes ane enigma configurations with AdS$_3$ asymptotics \cite{Gauntlett:2004wh,Denef:2007vg, deBoer:2008fk,Bena:2011zw}.

\subsubsection{Spontaneous symmetry breaking in the CFT side}
The Hawking-Page transition in the canonical ensemble between the thermal AdS and the large AdS black hole has long been understood as the confinement-deconfinement phase transition on the boundary CFT side \cite{Witten:1998zw,Klebanov:2000hb}
and can also be related to phase transitions in Randall-Sundrum brane-world scenarios \cite{Randall:1999ee,Randall:1999vf,Karch:2000ct} as recently analyzed in \cite{Mishra:2024ehr}. 

It would be interesting to understand the boundary mechanism that is responsible for the phase transition to the localized black holes in the microcanonical ensemble found in this work. In particular, the spontaneous breaking of the internal symmetry $SO(4)$ associated with the compact $\rm S^3$ to $U(1) \times U(1)$ is one of the most important features. 

When tracing the microcanonical ensemble by changing $E$, the most dominant solution changes from the global AdS$_3$ (at $E=-c/12$), to the black pole, then to the BTZ geometry, as shown in Fig.\ref{fig:NegEn}. While the $SO(4)$ internal symmetry is preserved in the global AdS$_3$ solution and the BTZ geometry, it is broken to $U(1)\times U(1)$ in the black pole. This indicates that in the strong-coupling limit of the D1-D5 CFT, there should be a spontaneous $SO(4)$ symmetry-breaking happening in the low-lying (but $\mathcal{O}(c)$) spectrum. At the lowest possible energy, the black pole disappears, and the internal symmetry becomes again  $SO(4)$.

A similar pattern of symmetry breaking occurs in the Klebanov-Strassler solution \cite{Klebanov:2000hb}. The high-energy phase is described by a black hole which preserves the chiral symmetry \cite{Aharony:2007vg}. This black hole dominates both the canonical and the microcanonical ensemble. As one lowers the energy, one finds new black-hole phases with broken chiral symmetry \cite{Buchel:2018bzp, Buchel:2024phy} and, as one further lowers the energy, one expects to find black holes localized on the $S^3$ at the tip of the KS throat \cite{Bena:2018fqc, Bena:2019sxm, Lust:2022xoq}, which break the isometries even further. When the energy is exactly zero, these localized black holes disappear, and the symmetry of the dominant phase is enhanced again. This behavior is different 
from most familiar examples of spontaneous symmetry breaking (SSB), such as the Ising model, where the symmetry breaking typically extends to the ground state. 

Moreover, while we have focused on specific asymptotically-$\rm AdS_3 \times S^3 \times T^4$ solution, we expect that SSB mediated by localized black holes occurs universally in asymptotically ${\rm AdS}_D \times \mathcal{M}$ spacetimes, as suggested by similar constructions in AdS$_5\times$S$^5$ \cite{Hubeny:2002xn,Buchel:2015gxa,Dias:2016eto}. Thus, SSB in the low-lying spectrum with $\mathcal{O}(1/G_N^{(D)})$ energy excitation could represent a novel characteristic of holographic CFTs.

\subsubsection{The eigenstate thermalization hypothesis with SSB}

While we still lack a clear understanding from the CFT perspective of why different black-hole geometries with the same energy are barely distinguishable through entanglement entropy, it is worth revisiting this question through the framework of the eigenstate thermalization hypothesis (ETH) \cite{Srednicki94, Deutsch:1991msp}.

The {\it strong ETH} asserts that\footnote{There are multiple versions and formulations of the ETH. The version considered here is known as the strong ETH or microscopic thermal equilibrium ETH. For a comprehensive review, see \cite{Mori2018}. For its connections to holography, see \cite{Kawamoto:2024vzd}.}, for any eigenstate $\ket{\psi_E}$ of a chaotic quantum many-body system with energy $E$, if the system is divided into two subsystems $A$ and $B$, and we focus on the smaller subsystem $|A| < |B|$, the reduced density matrix $\rho_A^{\psi} \equiv {\rm Tr}_B\left(|\psi\rangle\langle\psi|\right)$ can be well-approximated by the reduced density matrix of the microcanonical ensemble at the same energy:
\begin{align}\label{eq:strongETH}
    \rho_A^{\psi} \approx {\rm Tr}_B\left(\mathcal{N} \sum_{E' \in [E-\Delta E, E]} | E'\rangle \langle E'|\right) \equiv \rho_A^{\rm (micro)}. 
\end{align}
It follows directly that for any operator $O_A$ with its support on subsystem $A$, its expectation value evaluated on $\ket{\psi_E}$ and that evaluated on $\rho^{\rm micro}$ are approximately equal,
\begin{align}\label{eq:equalEV}
    \braket{\psi_E|O_A|\psi_E} \approx {\rm Tr} \left(\rho^{\rm (micro)} O_A\right),
\end{align}
and the entanglement entropy for subsystem $A$ is approximately equal in both $\ket{\psi_E}$ and $\rho^{\rm (micro)}$, i.e. 
\begin{align}\label{eq:equalEE}
    S_{\rm vN}\left(\rho_A^{\psi}\right) \approx S_{\rm vN}\left(\rho_A^{\rm (micro)}\right),
\end{align}
where $S_{\rm vN}$ stands for the von Neumann entropy. 

Our system does {\it not} satisfy the strong ETH in the following way. At a fixed energy within the range $c\left(5\sqrt{5}-11\right)/24 < E < c/96$, the black pole geometry and the BTZ geometry coexist, with $S_{\rm BP} < S_{\rm BTZ}$. There are $e^{S_{\rm BP}}$ microstates corresponding to the black pole and $e^{S_{\rm BTZ}}$ microstates corresponding to the BTZ black hole, and the microcanonical ensemble is dominated by the BTZ geometry. Consider an energy eigenstate described by the black pole geometry. This eigenstate admits a non-zero VEV for a simple CFT operator associated with the breaking of $SO(4)$ internal symmetry. In contrast, the microcanonical ensemble $\rho^{\rm (micro)}$, dominated by the BTZ geometry, preserves $SO(4)$ symmetry and exhibits a vanishing VEV for the same CFT operator. This discrepancy indicates that \eqref{eq:equalEV} is not satisfied with this choice of eigenstate. Therefore, our system does not satisfy the strong ETH.

If our system had satisfied the strong ETH, we would have been able to use it to derive \eqref{eq:equalEE} and explain why different black hole geometries with the same energy are nearly indistinguishable through entanglement entropy. In summary, while our system satisfies \eqref{eq:equalEE}, it fails to satisfy \eqref{eq:equalEV} and \eqref{eq:strongETH}.

Instead of attributing this feature to a coincidence, it is insightful to examine the current situation in greater detail. The microcanonical ensemble undergoes spontaneous symmetry breaking (SSB) at $E = c\left(5\sqrt{5}-11\right)$. Beyond this threshold, while the majority of microstates preserve the $SO(4)$ symmetry, as described by the BTZ geometry, a small fraction breaks the $SO(4)$ symmetry, as described by the black pole. The failure of \eqref{eq:equalEV} and \eqref{eq:strongETH} arises directly from this symmetry breaking.

Our analysis strongly suggests that this type of symmetry breaking, while altering the expectation values of a class of operators, does not significantly affect the entanglement structure. While systems with SSB cannot satisfy the conventional ETH, it is worth investigating the extent to which the universality of thermalization is preserved in chaotic quantum many-body systems with SSB. Generalizations of ETH to such systems have been discussed in \cite{fratus2015eigenstate}, but their framework is not robust enough to imply \eqref{eq:equalEE}.

Further exploration of the ETH in systems with SSB is an intriguing direction for future research. Our localized black holes could provide valuable insight into this line of inquiry.

\subsubsection{Holographic disentangled states and quantum scars}

Understanding the microstructure of black holes has been an ever-lasting question. Recently, tools from quantum information theory and statistical physics have provided new insights into this topic. In \cite{Hayden:2020vyo,Wei:2022cpj}, it was found that sufficiently many disentangled black hole microstates, with parametrically small entanglement entropy compared to typical microstates, exist to account for the leading order of the Bekenstein-Hawking entropy.\footnote{It was pointed out in \cite{Hayden:2020vyo} that the additivity conjectures in quantum information theory \cite{Shor2004}, which are possibly violated, imply the current statement. This statement was later shown in \cite{Wei:2022cpj} without assuming the additivity conjectures.}

A disentangled state is not necessarily an energy eigenstate. However, if it is, it qualifies as a {\it quantum scar} \cite{Bernien:2017ubn,Turner:2018kjz}, which is defined as an eigenstate that does not satisfy the strong ETH. While a quantum scar is not necessarily disentangled in principle, conventional examples of quantum scars in many-body systems typically have sub-volume law entanglement and are thus disentangled \cite{Serbyn:2020wys,Moudgalya:2021xlu}, in contrast to typical microstates that exhibit volume law entanglement.

By definition, an energy eigenstate described by the black pole geometry in the range $c\left(5\sqrt{5}-11\right)/24 < E < c/96$ is a quantum scar. This is because it can be distinguished from the microcanonical ensemble described by the BTZ geometry through a simple operator that acquire vacuum expectation values because of the $SO(4)$ symmetry breaking. However, this represents a ``boring" quantum scar, as its entanglement entropy is approximately the same as that of the microcanonical ensemble. This scar might be reconciled by considering a generalization of ETH with SSB, as discussed earlier. Other types of quantum scars in holographic CFT have been constructed, for example, in \cite{Dodelson:2022eiz,Milekhin:2023was}, but none of them are disentangled.

An intriguing open question is to identify disentangled quantum scars in AdS/CFT.\footnote{The analysis in \cite{Kusuki:2019rbk} suggests that certain high-energy primary states in AdS$_3$/CFT$_2$ might be disentangled quantum scars, exhibiting area-law entanglement in CFT computations, though their gravitational descriptions remain unclear.} It has been proposed in \cite{Hayden:2020vyo,Wei:2022cpj,Milekhin:2023was} that static horizonless geometries at high energy are promising candidates for disentangled quantum scars. Intuitively, this is because, while the microcanonical ensemble at high energy is dominated by large black holes (where the RT surface wraps the horizon, exhibiting volume-law entanglement), the RT surface in a horizonless geometry caps off smoothly in the bulk, likely exhibiting area-law entanglement.

Testing this intuition has been challenging because horizonless geometries with smooth capping often have nontrivial dependence on the compact directions. However, the techniques developed in Section \ref{sec:EEGen} of this paper enable the computation of holographic entanglement entropy in geometries with nontrivial fibrations over compact directions. We hope to leverage this technique to search for disentangled holographic quantum scars in the near future.

\section*{Acknowledgements}
We are grateful to Sougmangsu Chakraborty, Matthew Dodelson, Matthew Heydeman, Yuya Kusuki, Samir Mathur, Alexey Milekhin, Jorge Santos, Stephen Shenker, Douglas Stanford, Shreya Vardhan, Diandian Wang, and Yasushi Yoneta for useful discussions and comments. The work of IB and RD was supported in part by the ERC Grants 787320 ``QBH Structure'' and 772408 ``Stringlandscape.'' The work of PH is supported by the Department of Physics at The Ohio State University.
The work of ZW is supported by the Society of Fellows at Harvard University. IB would like to thank the Harvard Swampland Initiative for the hospitality that made possible the interactions that initiated this research project.

\appendix

\section{Entanglement entropy}
\label{app:1}

In this section, we detail the computation of holographic entanglement entropy in AdS$_3\times\cM$ \emph{static} spacetimes as a function of the boundary length, where $\cM$ is a compact space. We first review the derivation for three-dimensional, asymptotically AdS$_3$ spaces, before extending to ten-dimensional solutions asymptotic to AdS$_3\times$S$^3\times$T$^4$.

\subsection{Entanglement entropy in \texorpdfstring{AdS$_3$}{AdS3}}
\label{app:EEGen}

For static solutions, there is no ambiguity in selecting a time slice on which to derive the minimal surfaces. Each time slice is two-dimensional, and both the boundary and the Ryu-Takayanagi surface are one-dimensional. The holographic computation of the entanglement entropy is thus reduced to a geodesic problem. The corresponding boundary CFT of an asymptotically AdS$_3$ space lives on an S$^1$ of radius $2\pi$. We aim to compute the entanglement entropy between a subsystem $\mathcal{L}$ of this circle, where we denote the length of $\mathcal{L}$ by $L$. The RT formula \cite{RT06,RT06b} states that the entanglement entropy is given by 
\begin{equation}
    S_{EE}(L) \= \frac{1}{4G_3}\,\underset{\partial \gamma = \partial \cL}{\min} A(\gamma)\,, \qquad A(\gamma) \equi \int   ds_\gamma\,, 
\end{equation}
and $S_{EE}(L)$ is proportional to the length in AdS$_3$ of the geodesic connecting the boundary $\partial \mathcal{L}$ of $\mathcal{L}$ as a function of $L$.

As in the main text, we denote the AdS radial and angular directions by $r$ and $\phi$, respectively. The spacelike geodesics are governed by
\begin{equation}
    g_{rr}\left(\frac{\mathrm{d}r}{  d \lambda}\right)^2+g_{\phi\phi}\left(\frac{\mathrm{d}\phi}{ d \lambda}\right)^2=1\,,
\end{equation}
where $\lambda$ is an affine parameter. The geodesic length, $\cA\equi \underset{\partial \gamma = \partial \cL}{\min} A(\gamma)$, from which the entanglement entropy is derived, is then:
\begin{equation}
    \cA \=\int\mathrm{d}\lambda\,.
\end{equation}
Assuming that the geometries are independent of $\phi$, $\partial_\phi$ is a Killing vector, giving us the conserved quantity:
\begin{equation}
    \cE \=g_{\phi\phi}\, \frac{\mathrm{d}\phi}{ d \lambda}\,.
\end{equation}
This quantity parametrizes the minimal radius, $r_{\rm{min}}$, reached by the geodesic. Indeed, the equation of motion becomes: 
\begin{equation}
    g_{rr}\left(\frac{{ d}r}{{ d} \lambda}\right)^2+\frac{1}{g_{\phi\phi}} \,\cE^2=1\,.
\end{equation}
At $r=r_{\rm{min}}$, $\frac{dr}{ d \lambda}=0$, so $\cE^2=g_{\phi\phi}(r=r_{\rm{min}})$. We implicitly assume that $g_{\phi\phi}$ is a monotonic function of $r$, meaning that the larger $\cE$ is, the smaller $r_{\rm{min}}$ will be.

Finally, the boundary length, $L$, and the geodesic length, $\cA$, are given in terms of integrals from $r_{\rm min}$ to a UV cutoff, $r_\infty$:
\begin{align}
    &L \= \int {\rm d} \phi \= 2\int_{r_{\rm{min}}}^{r_{\infty}}  {\rm d} r\, \frac{\cE}{g_{\phi\phi}} \sqrt{g_{rr}\left(1-\frac{\cE^2}{g_{\phi\phi}}\right)}\,,\nn\\
    &\mathcal{A}\=\int {\rm d} \lambda \= 2\int_{r_{\rm{min}}}^{r_{\infty}} {\rm d}r \,\sqrt{g_{rr}\left(1-\frac{\cE^2}{g_{\phi\phi}}\right)}.
\end{align}
To express the geodesic length as a function of $L$, we should treat the first equation as an implicit definition of $r_{\rm{min}}$ and substitute it into the second equation.

\subsection{Entanglement entropy in \texorpdfstring{AdS$_3\times$S$^3\times$T$^4$}{AdS3 times S3 times T4}} 
\label{app:EELocBH}

In Section \ref{sec:RTForma}, we presented a method to adapt the RT formula for computing holographic entanglement entropy of ten-dimensional geometries asymptotic to AdS$_3 \times$ S$^3 \times$ T$^4$.

As explained in \cite{Giusto:2014aba},  when the metric along the AdS$_3$ coordinates is independent on the  S$^3$ coordinates, the area of can be done by reducing the problem to an effective three-dimensional geometry \eqref{eq:Eff3dMet}, which encapsulates the deformations of the S$^3 \times$ T$^4$ ``internal manifold'' as one moves inside AdS$_3$. 

However, our solutions are more complicated, and the RT surface depends non-trivially on $\theta$, the (Hopf) azimuthal angle on the S$^3$. To find its area we need to find a spacelike minimal curve $\gamma_0$ in the reduced three-dimensional space, evaluated at a fixed $\theta = \theta_0$. We then compute the ten-dimensional minimal area by evaluating $A(\gamma_0 \times S^3 \times T^4)$, \eqref{eq:MinSurfTheta0}, integrating over the S$^3 \times$ T$^4$ directions along $\gamma_0$. The entanglement entropy is finally obtained by minimizing over all possible angular positions $\theta_0$, as shown in \eqref{eq:EntanglementEntropLocBH}.

In this section, we provide more details on this derivation, applying it to our class of geometries, generally given by \eqref{eq:metAdS3+BTZs} and \eqref{eq:Eff3dMet}.

To determine the minimal curve $\gamma_0$, we define it as a spacelike geodesic in the effective three-dimensional metric \eqref{eq:Eff3dMet}, fixed at $\theta = \theta_0$ and constant time. The metric on this hypersurface is:
\begin{equation}
     ds^2\=Q_1Q_5 G(r,\theta_0)\, \left[ R_y^2\frac{r^2+\ell^2}{\cK_y(r,\theta_0)} d\phi^2 + Q_1Q_5G(r,\theta_0)\,\frac{ dr^2}{r^2+\ell^2}\right]\,.
\end{equation}
The geodesic equation becomes:
\begin{equation}
    Q_1Q_5\,R_y^2\,G(r,\theta_0)\frac{r^2+\ell^2}{\cK_y(r,\theta_0)} \left(\frac{d\phi}{ d\lambda}\right)^2 + (Q_1Q_5)^2\frac{G(r,\theta_0)^2}{r^2+\ell^2}\left(\frac{{d}r}{{d}\lambda}\right)^2=1\,,
\end{equation}
where $\lambda$ is the affine parameter. The $\phi$-isometry gives a conserved quantity:
\begin{equation}
    \cE\=Q_1Q_5\, R_y^2\,G(r,\theta_0)\frac{r^2+\ell^2}{\cK_y(r,\theta_0)} \left(\frac{{d}\phi}{{d}\lambda}\right)\,,
\end{equation}
leading to:
\begin{equation}
     \frac{(Q_1Q_5)^2G(r,\theta_0)^2}{r^2+\ell^2}\left(\frac{{d}r}{{d}\lambda}\right)^2=\left(1-\frac{\cE^2 \cK_y(r,\theta_0)}{ Q_1Q_5\,R_y^2\,G(r,\theta_0)\,(r^2+\ell^2)}\right)\,.
\end{equation}

We define the minimal radius reached by the geodesic, $r_{\mathrm{min}}$, as  $$\cE^2=\frac{Q_1Q_5\,R_y^2 \,G(r_{\mathrm{min}},\theta_0)\left(r_{\mathrm{min}}^2+ \ell^2\right)}{\cK_y(r_{\mathrm{min}},\theta_0)}.$$

The boundary length of $\gamma_0$ as a function of $r_{\rm min}$ and $\theta_0$ is given by:
\begin{equation}
    L(\gamma_0)\=\int {d}\phi=2\int_{r_{\mathrm{min}}}^{r_\infty} \frac{\cK_y(r,\theta_0)}{R_y(r^2+ \ell^2)}\, \frac{\sqrt{Q_1 Q_5\,(r_{\mathrm{min}}^2+ \ell^2)}\,d r}{\sqrt{(r^2+\ell^2)\,\frac{\cK_y(r_{\mathrm{min}},\theta_0)}{G(r_{\mathrm{min}},\theta_0)}-\left(r_{\mathrm{min}}^2+ \ell^2\right)\,\frac{\cK_y(r,\theta_0)}{G(r,\theta_0)}}}\,,\label{eq:BoundaryLengthApp}
\end{equation}
where $r_\infty$ is the UV cutoff, as in previous sections. This relation implicitly defines $r_{\rm min}(\theta_0)$ in terms of $L$. Using the geodesic equations, we derive the area of the codimension-one hypersurface along $\gamma_0$, $A(\gamma_0 \times S^3 \times T^4)$, as in \eqref{eq:MinSurfTheta0}:
\begin{align}
  A(\gamma_0 \times S^3 \times T^4) \= 2\,(2\pi)^6 \, V_4 \, \int &\,\sqrt{\scriptstyle{\left(\frac{G(r,\theta)}{G(r,\theta_0)}\right)^2 \left(1-\frac{(r_{\mathrm{min}}^2+ \ell^2)\,\cK_y(r,\theta_0)\,G(r_{\mathrm{min}},\theta_0)}{(r^2+ \ell^2)\,\cK_y(r_{\mathrm{min}},\theta_0)\,G(r,\theta_0)}\,\right)+\frac{G_{\phi\phi}(r,\theta)\,G_{\phi\phi}(r_{\rm{min}},\theta_0)}{G_{\phi\phi}(r,\theta_0)^2}}}\nonumber \\
  & \frac{Q_1 Q_5\,G(r,\theta_0)\,\cos \theta \sin \theta \,  d\theta\,{d}r}{\sqrt{(r^2+\ell^2)-\left(r_{\mathrm{min}}^2+ \ell^2\right)\,\frac{\cK_y(r,\theta_0) \,G(r_{\mathrm{min}},\theta_0)}{\cK_y(r_{\mathrm{min}},\theta_0)\,G(r,\theta_0)}}}\,, \label{eq:AreaEqApp}
\end{align}
where we introduced $G_{\phi\phi}(r,\theta)=\frac{(r^2+\ell^2)G(r,\theta)}{\mathcal{K}_y(r,\theta)}$. We then obtain the relationship $A(\gamma_0 \times S^3 \times T^4)$ as a function of $L(\gamma_0)$ and minimize over all angular positions $\theta_0$ to find the minimal area, as given in \eqref{eq:EntanglementEntropLocBH}.

\subsection{Small subsystems and the entanglement first law}
\label{app:SmallGeo}

In this section, we compute the entanglement entropy of our ten-dimensional geometries for small boundary lengths, $L$. This corresponds to small geodesics that remain close to the boundary. To achieve this, we expand the metric to the first non-trivial order in the radial distance $r$, using the generic expressions in \eqref{eq:FieldGeneric}:
\begin{align}
    \cK_y(\theta, r)=1+\sum_{i\in U_y}\frac{\ell_i^2}{ r^2}+\mathcal{O}(r^{-4})\,,\qquad G(r,\theta)=1+\mathcal{O}(r^{-4})\,.
\end{align}
Thus, for small geodesics, the S$^3$ fibration is trivial, as all dependence on $\theta$ drops out.

By implementing this large-$r$ expansion in \eqref{eq:BoundaryLengthApp}, we find that the boundary length $L$ is given in terms of $r_{\rm min}$ as
\begin{align}
    L &\=\frac{2 \sqrt{Q_1Q_5}}{R_{y}\,r_{\mathrm{min}}}\left(1-\frac{1}{r_{\mathrm{min}}^2}\left(\sum_{i\in U_y}\frac{\ell_i^2}{3}+\sum_{i\in U_t}\frac{\ell_i^2}{6}\right)\right),
\end{align}
which can be inverted to give:
\begin{equation}
    r_{\rm{min}}\=\frac{2\sqrt{Q_1Q_5}}{ L\,R_y}-\frac{ L\,R_y}{2\sqrt{Q_1Q_5}} \left(\sum_{i\in U_y}\frac{\ell_i^2}{3}+\sum_{i\in U_t}\frac{\ell_i^2}{6}\right)\,.
\end{equation}
Finally, the area in \eqref{eq:AreaEqApp} also integrates easily for small geodesics, giving: 
\begin{align}
    A(\gamma_0 \times S^3 \times T^4) \= (2\pi)^6 \, V_4\, Q_1Q_5 \left( \log\left(\frac{r_\infty}{r_{\mathrm{min}}}\left(1+\sqrt{1-\left(\frac{r_{\mathrm{min}}}{r_{\infty}}\right)^2}\right)\right)-\sum_{i\in U_y}\frac{\ell_i^2}{ 2\,r_{\rm{min}}^2}\right)\,. \nn
\end{align}
Expressing this in terms of the boundary length, $L$, yields: 
\begin{equation}
    A(\gamma_0 \times S^3 \times T^4)= (2\pi)^6 \, V_4\, Q_1Q_5 \left(\log\left(\frac{r_\infty  L\,R_y}{\sqrt{Q_1Q_5}}\right)+\frac{L^2 \,R_y^2}{4Q_1Q_5}\left(\sum_{i\in U_t}\frac{\ell_i^2}{ 6}-\sum_{i\in U_y}\frac{\ell_i^2}{ 6}\right)\right)\,.
\end{equation}
As expected, the area is independent of the angular position $\theta_0$ for small geodesics, so minimizing with respect to $\theta_0$ is trivial. Thus, we obtain the entanglement entropy \eqref{eq:ShortIntervalEE}.

\section{Black-hole pure states/mixed states and RT}\label{app:pure_mixed}

When analyzing the holographic entanglement entropy in the main text, we focused on pure states associated with black hole geometries.

In this Appendix, we explain how a black-hole geometry can correspond to different quantum states in the CFT, including both mixed and pure states, and how these distinctions are reflected in the RT computation. We provide general arguments and use the BTZ geometry as a specific example to illustrate these concepts.

\subsection{Black holes as canonical and microcanonical ensembles}

In this paper we analyzed the thermodynamics of the CFT dual to the AdS black hole geometry, both in the canonical and in the microcanonical ensemble.
This black hole describes the mixed state in both CFT ensembles, and this may seem puzzling, as it appears to contradict the one-to-one correspondence expected in AdS/CFT. However,  the fact that the same BTZ black hole describes different ensembles of CFT states is a consequence of the semiclassical approximation, where certain details have been dropped. 

To clarify this puzzle, let us focus on a BTZ black hole geometry with energy $E > c/12$ and temperature $T > 1/2\pi$. On one hand, this black hole  solution corresponds to the canonical mixed state at temperature $T$:
\begin{align}
    \rho^{\rm (can)} = \frac{e^{-H_{\rm CFT}/T}}{{\rm Tr}\left(e^{-H_{\rm CFT}/T}\right)}.
\end{align}
This connection can be understood as follows. On the CFT side, the canonical mixed state is constructed via a standard Euclidean path integral on a torus $\rm T^2$. Using the Gubser-Klebanov-Polyakov-Witten (GKPW) dictionary \cite{GKP98,Witten98}, this corresponds to a gravitational path integral in the bulk AdS$_3$, summing over all bulk geometries with the same $\rm T^2$ boundary. In the semiclassical limit ($G_N \rightarrow 0$ or equivalently $c \rightarrow \infty$), the saddle-point approximation becomes applicable, reducing the problem to finding the bulk solution with minimal action (or equivalently, minimal free energy). The BTZ black hole geometry serves as this saddle point, thus providing a semiclassical approximation to the canonical Gibbs state in the CFT.

On the other hand, the BTZ geometry also corresponds to a microcanonical mixed state at energy $E$ described by the density matrix
\begin{align}
    \rho^{\rm (micro)} \= \mathcal{N} \sum_{E_i \in [E-\Delta E, E]} | E_i\rangle \langle E_i|, 
\end{align}
where $\mathcal{N}$ is a normalization factor. This correspondence arises from considering the microcanonical path integral \cite{Marolf18}, which prepares the microcanonical ensemble on the CFT side. In the semiclassical limit, the bulk problem reduces to finding the configuration that maximizes an alternative functional, which coincides with the Bekenstein-Hawking entropy for black hole geometries. The resulting configuration is again the BTZ black hole. For further details, see \cite{Marolf18}.

Thus, the BTZ geometry provides a semiclassical approximation to both the canonical and microcanonical ensembles on the CFT side. Although these two ensembles represent distinct mixed states in the CFT, they are characterized by the same macroscopic variables and are therefore described by the same BTZ geometry at leading order in the semiclassical (large-$c$) limit. This reflects the equivalence of thermodynamic ensembles in AdS/CFT. However, the differences between the two ensembles persist in the details of the gravitational path integral.

\begin{figure}
    \centering
    \includegraphics[width=16cm]{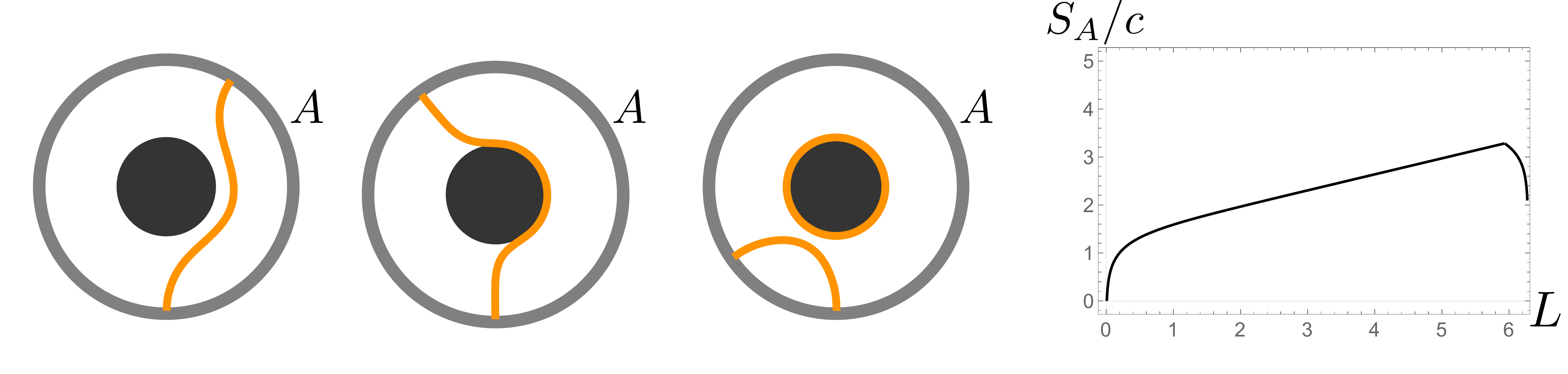}
    \caption{The left three figures show the minimal RT surfaces (orange curve) as a function of the length $L$ of subsystem $A$, when considering mixed states associated with the BTZ geometry. When $A$ becomes large, the RT surface includes the black hole horizon due to the homology condition. The right plot shows the holographic entanglement entropy as a function of $L$, with $E = c/3$ and $\epsilon = 0.01$. }
    \label{fig:HEE_mixed_states}
\end{figure}

When considering these mixed states (in either the canonical or the microcanonical ensembles associated with a given black hole geometry) and computing the holographic entanglement entropy using the RT formula, the black-hole horizon is treated as having nontrivial homology. Specifically, the bulk region enclosed by the subsystem $A$ and its RT surface must exclude the black hole \cite{RT06,RT06b}. See Fig.~\ref{fig:HEE_mixed_states} for an illustration.

As a result, for a BTZ geometry with energy $E$, the holographic entanglement entropy is given by:
\begin{equation}
 S_{EE\,BTZ}^{\rm mixed~state}(L) \= \min \begin{cases}
\frac{c}{3} \log\left[\sqrt{\frac{c}{3E}}\,\frac{1}{\epsilon} \,\sinh \left(\sqrt{\frac{3E}{c}}\,L \right) \right]\\
\frac{c}{3} \log\left[\sqrt{\frac{c}{3E}}\,\frac{1}{\epsilon} \,\sinh \left(\sqrt{\frac{3E}{c}}\,(2\pi -L) \right) \right] + 2\pi  \sqrt{\frac{cE}{3}}. 
\end{cases}
\label{eq:mixed_state}
\end{equation}
The right panel of Fig.\ref{fig:HEE_mixed_states} shows a plot of the holographic entanglement entropy as a function of $L$. Notably, $S_{EE}(L) \neq S_{EE}(2\pi - L)$, as we are dealing with mixed states in this context.

\subsection{Black holes as pure-state representatives}

A black hole geometry can also be interpreted as describing the macroscopic features of pure states on the CFT side. Given a black hole geometry with entropy $S$, one can associate a sub-Hilbert space with dimension $e^S$. A typical pure state $\ket{\psi}$ in this sub-Hilbert space can be approximated by the black hole geometry.

The difference between two such pure states can be reflected by the details of the geometry at or behind the event horizon. From a CFT perspective, examples of such pure states include late-time states in thermalization dynamics as discussed in\cite{Takayanagi:2010wp,Nozaki:2013wia,Hartman:2013qma,Asplund:2014coa,Ugajin:2013xxa,Shimaji:2018czt,Caputa:2019avh,Akal:2020twv,Akal:2021foz,Akal:2022qei,Miyata:2024gvr,Bai:2024azk}, or eigenstates satisfying the eigenstate thermalization hypothesis (ETH) \cite{Srednicki94,Mori2018}, as explored in \cite{Hartman:2013mia,Fitzpatrick:2014vua,Fitzpatrick:2015zha}. 

If we describe the pure state using the black hole geometry, 
the computation of the holographic entanglement entropy must assume that the horizon has trivial homology (see for example \cite{Takayanagi:2010wp,Nozaki:2013wia,Hartman:2013qma,Asplund:2014coa}).\footnote{For pure states admitting horizonless bulk duals, we do not encounter the issue for determining the homology condition.} 

\begin{figure}[t]
    \centering
    \includegraphics[width=16cm]{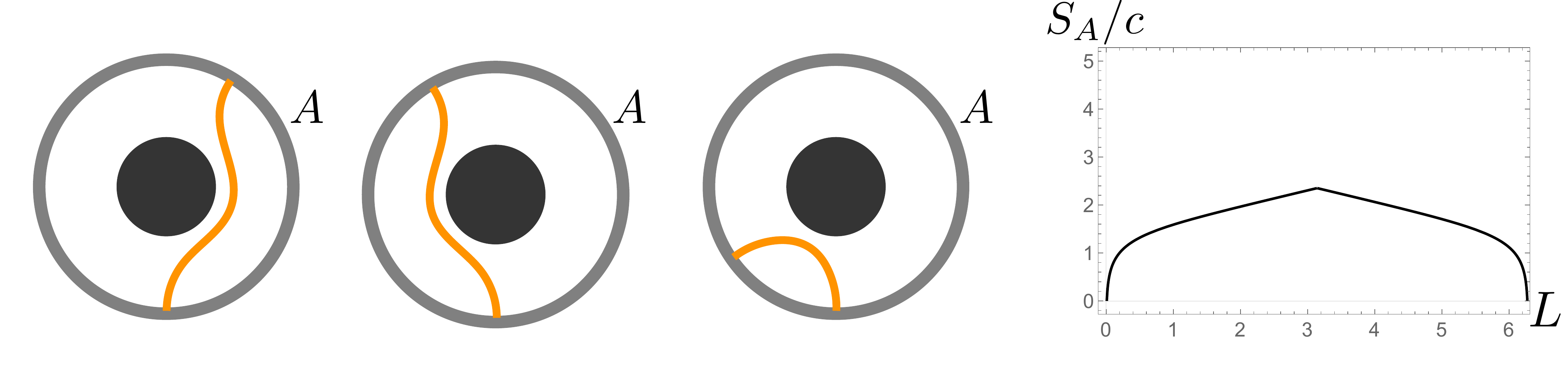}
    \caption{The left three figures show how the minimal RT surfaces (orange curve) change as a function of the length $L$ of the subsystem $A$ for a pure state associated with the BTZ geometry. The RT surface can pass through the black-hole region due to the trivial homology condition. The right plot shows the holographic entanglement entropy as a function of $L$, with parameters set to $E = c/3$ and $\epsilon = 0.01$.}
    \label{fig:HEE_pure_states}
\end{figure}

This means the bulk region surrounding subsystem $A$ and its RT surface may include the black hole itself. See Fig.\ref{fig:HEE_pure_states} for a sketch. For the BTZ geometry with energy $E$, the holographic entanglement entropy is:
\begin{equation}
 S_{EE\,BTZ}^{\rm pure~state}(L) \= \begin{cases}
\frac{c}{3} \log\left[\sqrt{\frac{c}{3E}}\,\frac{1}{\epsilon} \,\sinh \left(\sqrt{\frac{3E}{c}}\,L \right) \right]\\
\frac{c}{3} \log\left[\sqrt{\frac{c}{3E}}\,\frac{1}{\epsilon} \,\sinh \left(\sqrt{\frac{3E}{c}}\,(2\pi -L) \right) \right] , 
\end{cases}
\label{eq:pure_state}
\end{equation}
as shown in Fig.\ref{fig:HEE_pure_states}.

In the main text, we consider {\it pure states (microstates)} associated with the black-hole geometries. The entanglement entanglement entropy for pure states is symmetric, as shown in \eqref{eq:symmEnt}:
\begin{align}
    S_{EE}(L) = S_{EE}(2\pi-L).
\end{align}
Therefore, it is sufficient to consider $L \leq \pi$ for the calculation of entanglement entropy.

\subsection{Thermofield double states}

We have discussed how the treatment of the RT-surface homology condition differs depending on whether we consider the black hole as describing mixed states or pure states. One might wonder about the microscopic origin of this distinction. To address this, it is useful to consider {\it purifications} of mixed states. Given a mixed state $\rho$ in the Hilbert space $\mathcal{H}$, a purification of it is a pure state in an enlarged Hilbert space $\mathcal{H} \otimes \mathcal{H'}$ that recovers $\rho$ after tracing out the $\mathcal{H}'$ part, i.e.,
\begin{align}
    \rho = {\rm Tr}_{\mathcal{H}'} |\psi\rangle\langle\psi|. 
\end{align}
The advantage of considering the purification of a mixed state is that it allows us to treat the state with the same homology condition as pure states.

Let us now consider the canonical Gibbs mixed state,
\begin{align}
    \rho^{\rm (can)} \propto e^{-H_{\rm CFT}/T} = \sum_i e^{-E_i/T} |E_i\rangle \langle E_i|,
\end{align}
which corresponds to a BTZ geometry. This state admits a canonical choice of purification known as the thermofield double (TFD) state,
\begin{align}
    |{\rm TFD}\rangle \propto \sum_i e^{-E_i/2T}|E_i\rangle|E_i\rangle',  
\end{align}
which is defined on a doubly coupled Hilbert space $\mathcal{H} \otimes \mathcal{H'}$, where $\mathcal{H}'$ is isomorphic to the original Hilbert space $\mathcal{H}$. By tracing out the $\mathcal{H}'$ part, we recover the canonical ensemble:
\begin{align}
    {\rm Tr}_{\mathcal{H}'}\left(|{\rm TFD}\rangle\langle{\rm TFD}|\right) \propto \sum_i e^{-E_i/T} |E_i\rangle \langle E_i| \propto \rho^{\rm (can)}. 
\end{align}
The gravity dual of the TFD state is given by the time-reflection symmetric time slice of a two-sided BTZ black hole. If we take an interval on one side and look for its corresponding RT surface in the bulk, then one gets the picture depicted in Fig.\ref{fig:HEE_mixed_states} when viewed from one side.

\bibliographystyle{jhep}
\bibliography{BubbleEE}

\end{document}